\documentclass[aps,pre,twocolumn,superscriptaddress,floatfix,10pt]{revtex4}
\usepackage{graphicx}
\usepackage{dcolumn}
\usepackage{bm}
\usepackage{latexsym}
\usepackage{amsmath}
\usepackage{amssymb}
\usepackage{color}

\begin{document}

\title{A Biologically Motivated Asymmetric Exclusion Process: interplay of \\congestion in RNA polymerase traffic and slippage of nascent transcript {\footnote{The first two authors contributed equally}}}
 \author{Soumendu Ghosh}
\affiliation{Department of Physics, Indian Institute of Technology
  Kanpur, 208016, India}
\author{Annwesha Dutta}
\affiliation{Department of Physics, Indian Institute of Technology
  Kanpur, 208016, India}
\author{Shubhadeep Patra{\footnote{Undergraduate Visitor at IIT Kanpur}}} 
\affiliation{ISERC, Visva-Bharati, Santiniketan 731235}
\author{Jun Sato}
\affiliation{Research Center for Advanced Science and Technology, The University of Tokyo, Komaba 4-6-1, 
Meguro-ku, Tokyo 153-8904, Japan}
\author{Katsuhiro Nishinari}
\affiliation{Research Center for Advanced Science and Technology, The University of Tokyo, Komaba 4-6-1, 
Meguro-ku, Tokyo 153-8904, Japan}
\author{Debashish Chowdhury{\footnote{Corresponding author; e-mail: debch@iitk.ac.in}}}
\affiliation{Department of Physics, Indian Institute of Technology
  Kanpur, 208016, India}
  
\begin{abstract}
We develope a theoretical framework, based on exclusion process, that is motivated by a biological phenomenon called transcript slippage (TS). In this model a discrete lattice repesents a DNA strand while each of the particles that  hop on it unidirectionally, from site to site, represents a RNA polymerase (RNAP). While walking like a molecular motor along a DNA track in a step-by-step manner, a RNAP simultaneously synthesizes a RNA chain; in each forward step it elongates the nascent RNA molecule by one unit, using the DNA track also as the template. At some special ``slippery'' position on the DNA, which we represent as a defect on the lattice, a  RNAP can lose its grip on the nascent RNA and the latter's consequent slippage results in a final product that is either longer or shorter than the corresponding DNA template. We develope an exclusion model for RNAP traffic where the kinetics of the system at the defect site captures key features of TS events. We demonstrate the interplay of the crowding of RNAPs and TS.
A RNAP has to wait at the defect site for longer period in a more congested RNAP traffic, thereby increasing the likelihood of its suffering a larger number of TS events. The qualitative trends of some of our results for a simple special case of our model are consistent with experimental observations. The general theoretical framework presented here will be useful for guiding future experimental queries and for analysis of the experimental data with more detailed versions of the same model.
\end{abstract}

\maketitle
\section{Introduction}

Each single DNA strand is a hetero-polymer whose monomeric subunits are called nucleotides. By convention, the letters `A', `T', `C', and `G', represent the four nucleotide bases of a DNA. The specific sequence in which  these four letters appear in a DNA strand is a chemically-encoded genetic message (genetic information). Transcription of the genetic message is carried out by a  molecular machine called RNA polymerase (RNAP) \cite{buc09}. A RNAP synthesizes a RNA molecule (the transcript) whose sequence of monomeric subunits is complementary to that of a specific segment of its DNA that encodes the corresponding genetic message. The four letters of the alphabet that store genetic messages in the RNA transcript are `A', `U', `C', `G'. Each RNAP can also be regarded as a molecular motor \cite{chowdhury13a,chowdhury13b,kolomeisky15} for which the DNA template also serves as the track for its unidirectional, albeit noisy, movement during transcription \cite{buc09}. In each step forward, by one unit, along its DNA track, the RNAP elongates the nascent RNA transcript by one unit where unit is measured in terms of a nucleotide.

Experiments revealed the existence of specific stretches of DNA sequence where the nascent RNA may slip,  backward or forward, with respect to the RNAP although the RNAP motor itself does not slip simultaneously on its DNA track \cite{atkins10}. In fact, at any given slippage-prone site, multiple successive events of backward and/or forward slippage may occur before the correct transcription can resume. 
Thus, transcript slippage (TS) results in length heterogeneity of the final products of transcription because of the incorporation of more or fewer nucleotides, respectively, as compared to the length of transcript encoded in the DNA. While this phenomenon has received much attention over the past few decades \cite{atkins10,turnbough11},  the detailed mechanism of TS, its causes and consequences are still unclear. In this paper we focus on the consequences, rather than the causes, of TS.

Often the same segment of DNA (here loosely defined as the `gene') is simultaneously transcribed by several RNAPs, each synthesizing a distinct copy of the same RNA. The collective movement of multiple RNAPs simultaneously on a DNA track resembles, at least superficially, vehicular traffic on highways \cite{schadschneider10,chowdhury00}.
Wide varieties of collective traffic-like phenomena in non-living as well as in living systems have been modelled by various appropriate extensions of the totally asymmetric simple exclusion process (TASEP) \cite{schadschneider10,chowdhury00,chowdhury05,chou11,rolland15}.
In the past, RNAP traffic has been modelled theoretically \cite{tripathi08,klumpp08,klumpp11,sahoo11,ohta11,wang14,belitsky18} by extending the TASEP \cite{derrida98,schutz00,schadschneider10,mallick15}. 
A RNAP is expected to dwell longer at the slippery site in congested traffic because of the hindrance caused by the leading vehicle. The longer a RNAP dwells at the slippery site, the larger is the number of TS events it is likely to suffer. Thus, traffic congestion can influence the extent of TS. 
Here we develop a TASEP-based model to investigate the interplay of TS and RNAP traffic.

The  TASEP can never be in thermodynamic equilibrium, but can attain non-equilibrium steady states (NESS) \cite{derrida98,evans05,blythe07}. One of the key properties of the NESS is the non-vanishing particle flux which is defined as the number of particles passing through a site per unit time. The effects of different types of defects and inhomogeneities on the flux and the density profile of the particles have been investigated extensively over the last three decades \cite{janowski94,tripathy97,tripathy98,barma06,kolomeisky98,harris04,juhasz06,foulaadvand08,greulich08a,greulich08b,dong12,schmidt15,dhiman16,dhiman17,mishra17}. 
We treat the `slippage prone site', where TS occurs, as a special type of `defect' in a TASEP-based model of RNAP traffic.

In traffic engineering it is essential to first characterize the driving behaviors of individual drivers before embarking on a study of vehicular traffic on highways. In the same spirit, here we study the statistical characteristic of a single RNAP that undergoes TS at a specific location on the DNA track, before studying of RNAP traffic on the same track.
Thus, in this paper, we consider two different situations. In the first, a single RNAP is assumed to be moving alone on the DNA track whereas, in the second, many RNAPs move simultaneously in the same direction on a single DNA track. More specifically, we study three aspects of TS:  (a) how is the rate of transcription by a single RNAP affected by TS?, (b) how the error due to TS is affected by the traffic-congestion during the collective movement of RNAPs? and (c) how, in turn, the collective traffic-like movement of RNAP is affected by TS statistics? 

To study the effect of a transcript slippage on the movement of each individual RNAP on the DNA track and, equivalently, the rate of transcription, we use the technique of calculating First Passage Time (FPT) \cite{redner01,redner14,biswas16}. With this objective, we first construct a stochastic kinetic theory
that incorporates the effect of arbitrary numbers of backward and forward slippage. Then, as a concrete example, we consider a special case of the model that can be treated analytically without much of mathematical difficulty. For this concrete case, we compute the time taken by the RNAP motor to traverse a slippery site {\it for the first time}. This time is intrinsically stochastic and is termed here as the first-passage time. We interprete the results physically to explain how the movement of a RNAP on the DNA template (track) is affected by TS.

To study the interplay of RNAP traffic and TS, using Mean Field Approximation (MFA), we again compute the mean time needed to traverse a slippage site in a traffic of RNAPs on the same DNA track. We also carry out  Monte Carlo simulations of the model and compare the  Monte Carlo simulation data with the corresponding mean-field theoretic predictions to test the level of accuracy of the MFA. Finally, we also compare the theoretically predicted probability distribution of the longer and shorter transcripts with the experimental data \cite{olspert15,olspert16} obtained through advanced sequencing technologies  \cite{beverly18}.

The paper is organized in the following manner. In sec. II, we begin by sketching a brief introduction to the phenomenon of TS, followed by the description of our stochastic kinetic model of TS. In sec. III, we study the effect of TS on a single RNAP. More specifically, we derive an exact analytical expression for the mean time taken by a single RNAP, in the absence of steric hindrance from any other RNAP, to traverse the site where TS is likely to occur. In sec. IV, we investigate the effect of RNAP traffic congestion on TS. 
Our results establish that, because of traffic congestion, on the average, each RNAP dwells for longer time at the defect site and hence suffers a larger number of TS events. In sec. V, we analyze the effect of TS on RNAP traffic flow; the average density profile exhibits features that are typical characteristics of the TASEP with point defects. Finally, in sec. VI, we present a summary of the results and draw conclusions.  

\section{Model and Biological Motivation}

We begin this section with a brief overview of the TS process, as depicted schematically in  
Fig.\ref{fig-slippage_model}, in the subsection \ref{subsec-TSphenomenon}. Then, motivated 
by this biological phenomenon, in the subsection \ref{subsec-TSmodel} we develop our 
theoretical model. The distinct kinetic states are displayed, and the inter-state transitions are 
indicated, in Fig. \ref{fig-FP_model}.

\subsection{TS phenomenon}
\label{subsec-TSphenomenon}

\begin{figure}[h]
(a)\\[0.05cm]
\includegraphics[angle=0,width=0.45\columnwidth]{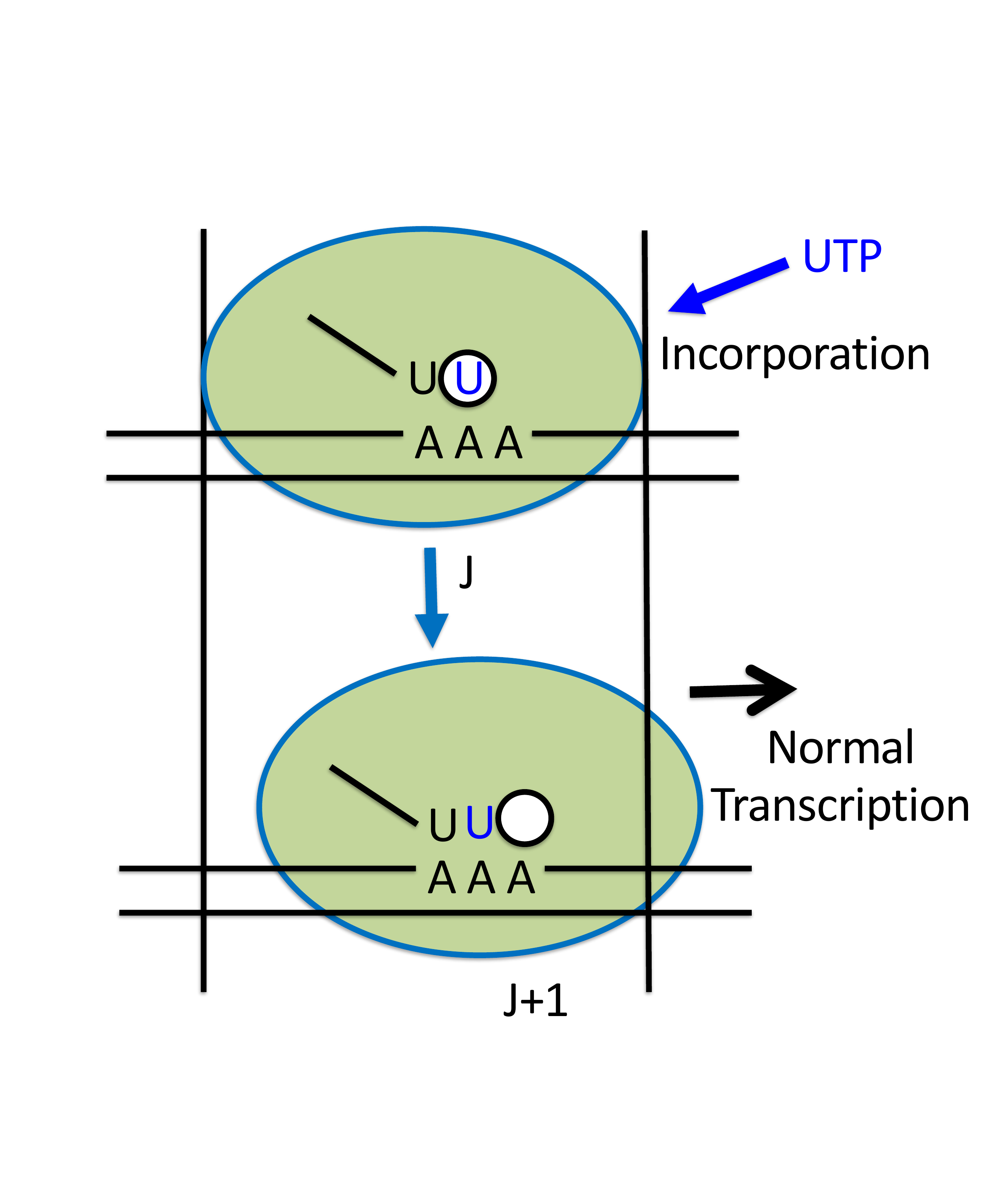}\\[0.05cm]
(b)~~~~~~~~~~~~~~~~~~~~~~~~~~~~~~~~(c)\\[0.05cm]
\includegraphics[angle=0,width=0.45\columnwidth]{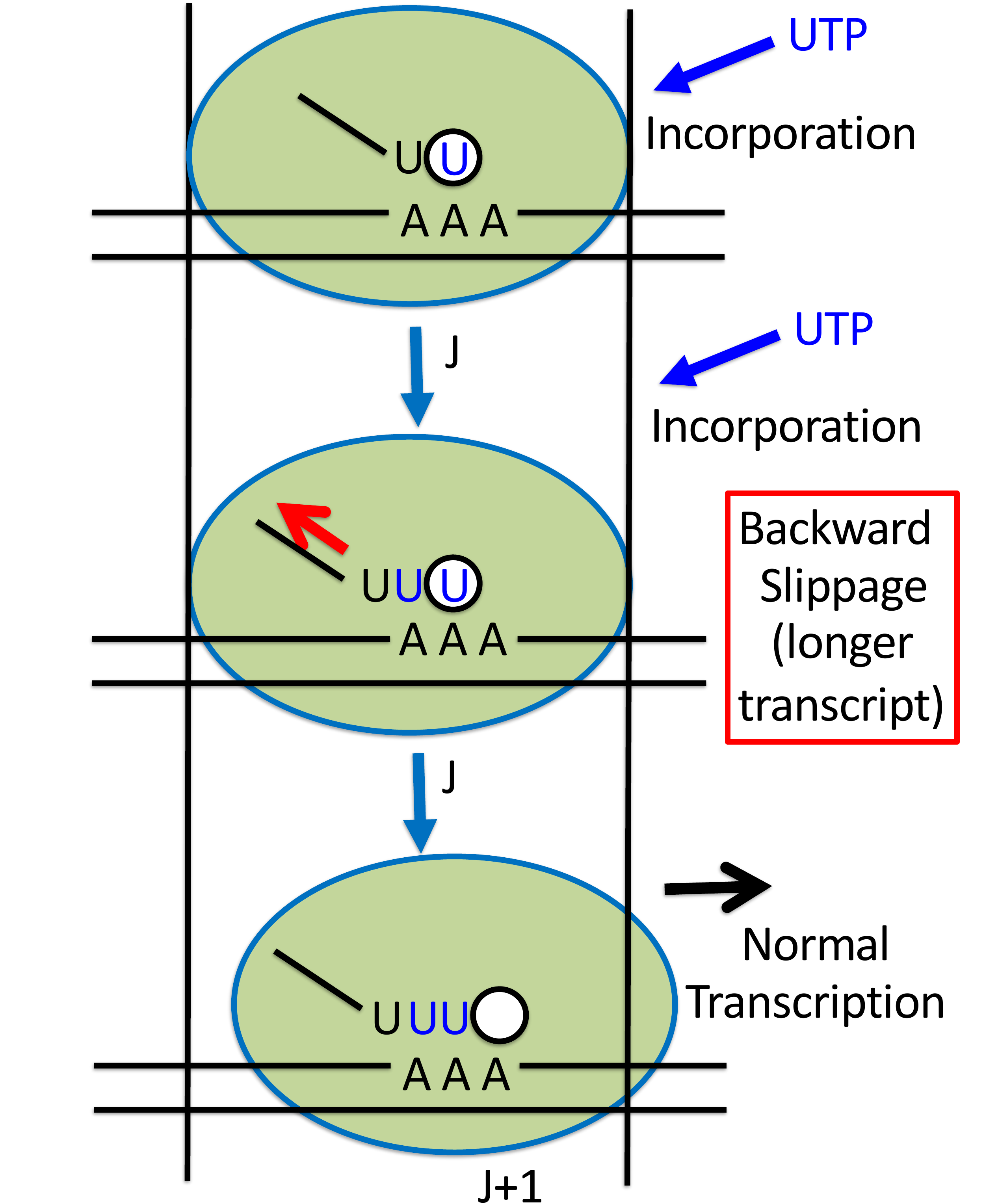}
\includegraphics[angle=0,width=0.45\columnwidth]{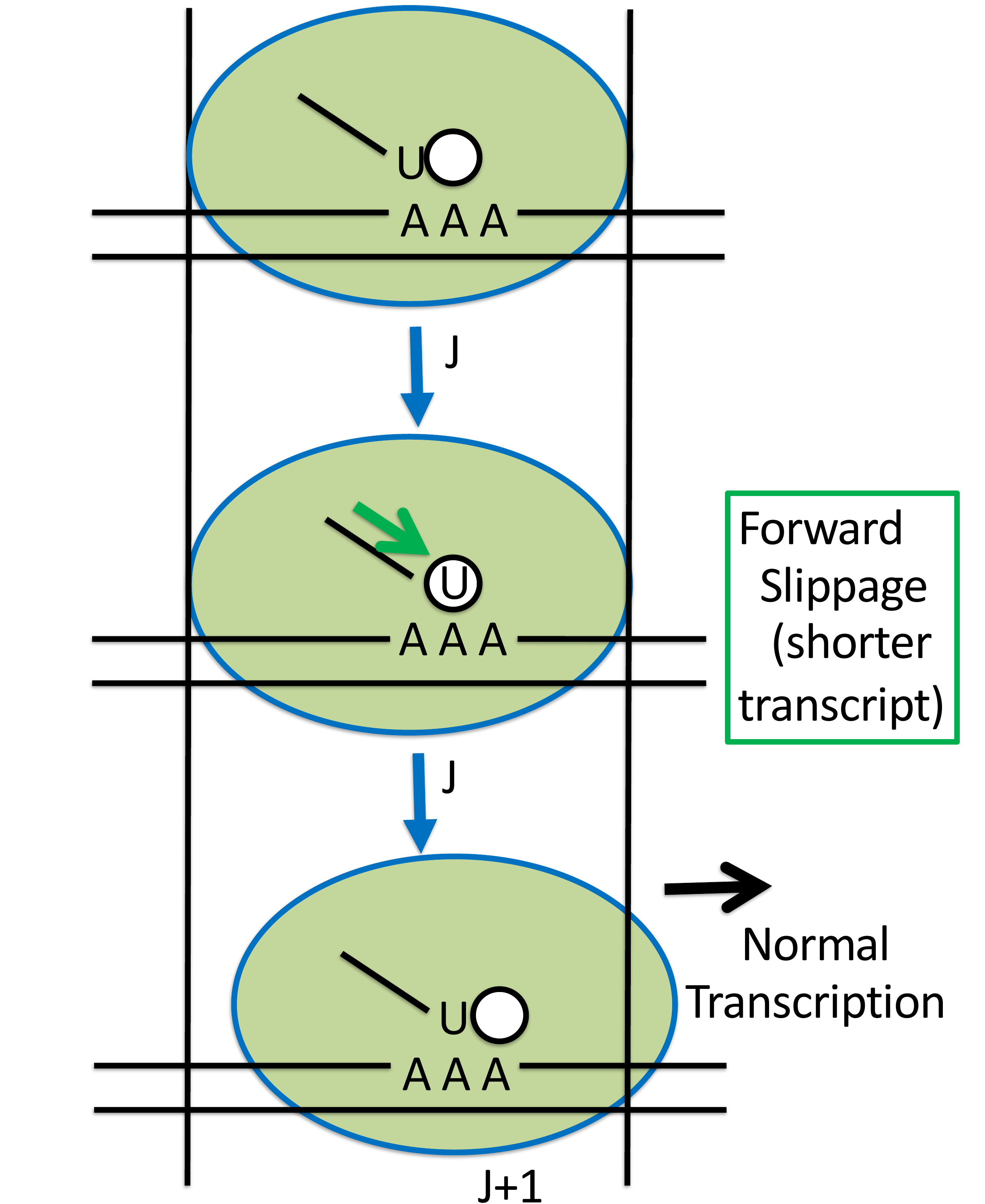}\\[0.05cm]
\caption{A pictorial depiction of (a) normal transcription, without slippage, by an RNAP, (b) transcription with backward slippage of nascent RNA, (c) transcription with forward slippage of nascent RNA. The white circle represents active site of an RNAP and blue color letter `$U$' depicts the incorporation of a nucleotide in the active site. Array of nucleotides `$A$' represents slippery sequence in the DNA strand and the slanted black solid line represents nascent RNA. Backward and forward slippages of the nascent RNA transcript are indicated by red and green arrows, respectively. In Fig (a), after incorporation of a nucleotide  `$U$', RNAP can move one step forward with respect to DNA template. In Fig (b), after incorporation of a nucleotide `$U$', nascent RNA slips backward with respect to RNAP as well as DNA template, by keeping RNAP fixed in its position. This results in addition of an extra nucleotide `U' to the transcript. In Fig (c), before the incorporation of a `$U$', nascent RNA can slip forward with respect to RNAP as well as DNA template, by keeping RNAP fixed in its position, resulting in a shortening of the transcript by one nucleotide. }
\label{fig-slippage_model}
\end{figure}

During normal transcription (see Fig.\ref{fig-slippage_model} (a)), incorporation of every nucleotide, as directed by the template DNA, is followed by the motor-like forward movement of the RNAP by one nucleotide along the DNA that also serves as its track. Thus, in normal transcription, each event of elongation of the nascent RNA transcript by one nucleotide is tightly coupled to the translocation of RNAP by one nucleotide on the template DNA. A common slippery sequence on the DNA is a sequence of A's  \cite{atkins10} (recall `A' is one of the four different types of subunits of DNA).  Fig.\ref{fig-slippage_model} (b) describes a transcription process wheere a single backward slippage of the nascent RNA transcript occurs. Due to the backward slippage, the active site of the RNAP turns empty for the second time during its sojourn at the slipper site, and it transcribes the same nucleotide for a second time, resulting in the insertion of an extra nucleotide on the transcript. Fig.\ref{fig-slippage_model} (c) describes a single forward slippage of the nascent RNA transcript without forward movement of RNAP; this forward slippage of the nascent transcript results in the active site of the RNAP getting occupied with the previously added nucleotide in the transcript thereby preventing incorporation of a fresh nucleotide even though the RNAP moves forward by one nucleotide on the DNA template. 

It is worth emphasizing that the slippage phenomena described above do not imply any movement of the RNAP with respect to its DNA track. Instead, in this paper we are concerned with the slippage of nascent RNA with respect to the RNAP as well as the DNA template.  In both Fig. \ref{fig-slippage_model} (b) and (c), the transcript slips without the concomitant movement of RNAP and that the polymerization of the transcript is not coupled with the translocation of RNAP. Typically, either of these mechanisms can be repeated thereby causing multiple rounds of backward or forward slippage of the transcript at these slippery sites.

\subsection{Kinetic model motivated by TS}
\label{subsec-TSmodel}

In our model, we represent the DNA template as a one-dimensional lattice of length $L$. We label the sites of the lattice by the integer index $j$ ($1 \leq j \leq L$). Each lattice site corresponds to a nucleotide on the DNA template. The instantaneous position of a RNAP is denoted by the integer index $j$; in each round of successful error-free elongation of the nascent RNA by one unit, the RNAP takes a forward step from $j$ to $j+1$. The special site where TS can take place has been labelled by the integer $J$ (i.e., $j=J$). Since our study is primarily on TS, and since TS is known \cite{atkins10} to occur at a special slippery site, we focus in this section exclusively on the triplet of sites $J-1$, $J$ and $J+1$.

For a completely normal error free transcription of the full length template DNA, the RNAP takes $L$ steps on the track synthesizing a RNA transcript of length $L$ i.e., exactly equal to the length of the DNA template. However, in case of transcription with $n$ successive rounds of backward slippage at a specially designated slippery site,  insertion of $n$ number of nucleotides leads to the synthesis of a longer transcript of total length $L+n$. Similarly, for $n$ successive rounds of forward slippage at the slippery site, missing the transcription of $n$ nucleotides on the templete (i.e., effectively,  deletion of $n$ nucleotides) produces a shorter  transcript of total length $L-n$. 
Backward slippages have been found to occur more often than the forward slippage. 

The extra length of the nascent RNA caused by the slippage is labelled by an integer index $\mu$ that can, in principle, be positive, negative or zero. According to our convention $\mu$ is positive (negative) in case backward (forward) slippage; in contrast,  $\mu =0$ if the nascent transcript suffers no slippage or it suffers equal numbers of forward and backward slippages at the slippery site $J$. In other words $\mu$ denotes the slippage-induced length change of the product transcript as compared to that of the template. Throughout this paper we use the term ``{\it slippage state}'' to denote the magnitude of $\mu$. 

\begin{figure}[h]
\begin{center}
\includegraphics[angle=0,width=0.95\columnwidth]{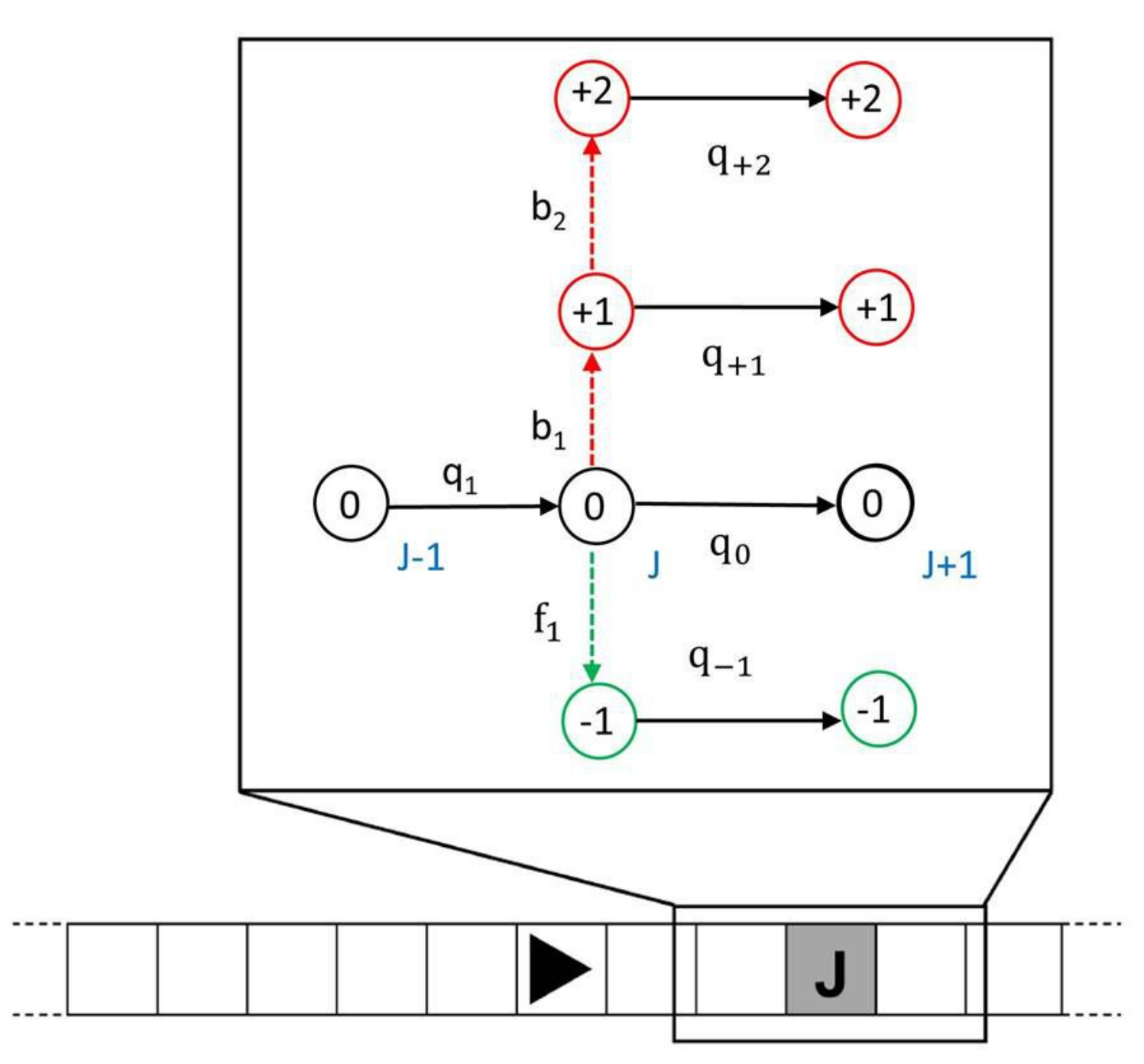}\\[0.02cm]
\end{center}
\caption{A kinetic model for the transcription slippage at site $J$ on the DNA template. The conformational states of RNAP during  transcription with first backward slippage of nascent mRNA transcript, transcription with second backward slippage of nascent mRNA transcript, transcription with forward slippage of nascent mRNA transcript and normal transcription are represented by +1, +2, -1 and 0 respectively. $b_{1}$, $b_{2}$ and $f_{1}$ are the first backward slippage rate (red arrow), second backward slippage rate (red arrow) and first forward slippage rate (green arrow), respectively. $q, q_0, q_{+1}, q_{-1}$ and $q_{+2}$ are the normal transcription rates (black arrows).}
\label{fig-FP_model}
\end{figure}

The theoretical framework that we have formulated is very general and can treat any arbitrary number $N_{b}$ of backward or $N_{f}$ number of forward slippage of the transcript while the RNAP is occupying the specific lattice site $J$. However, for the purpose of presentation of concrete results here through an explicit analytical calculation, we have allowed a maximum of two backward slippage events ($N_{b}=2$,  corresponding to $\mu= +1, +2$),  and a maximum of a single forward slippage ($N_{f}=1$ that would correspond to $\mu=-1$). Since backward slippages have been found to occur more often than the reverse process, in the example shown in Fig. \ref{fig-FP_model}, possibilties of two successive backward slippage  are shown against the possibility of a single forward slippage. 
For the same reason, we exclude the possibility of a backward slippage followed by a forward slippage which could result in a RNA of normal size. Moreover, allowing larger number of slippage (i.e., larger value of $|\mu|$) would create jam in RNAP traffic because at the slippery site each RNAP would have to wait much longer because of the larger number of slippage of its nascent mRNA due to higher allowed value of $|\mu|$.

The allowed transitions are indicated by the arrows and the corresponding rates are also shown next to the respective arrows in Fig. \ref{fig-FP_model}. The special case shown in Fig. \ref{fig-FP_model} is, indeed, simple enough to be treated analytically. However,  analytical calculations become more and more difficult with the increasing number of backward (or forward) slippage events (i.e., with the increase in the allowed values of $N_{b}$ and $N_{f}$). Nevertheless, in principle, the strategy of modelling followed here can be implemented numerically also for any arbitrary values of $N_{b}$ and $N_{f}$ if calculations becomes too difficult to carry out analytically.

Normal transcription, without any slippage, at the special site $J$ would correspond to the transition $(J,0) \rightarrow (J+1,0)$. In contrast, a backward TS at the site $J$ results in the transition $(J,0) \rightarrow (J,+1)$. Therefore, a single backward TS followed by a normal transcription together result in the composite transition  $(J,0) \rightarrow (J+1,+1)$. Since no TS are allowed to occur at any site before or after the special site $J$, the value of $\mu$ attained finally at $J$ is carried througout the subsequent transit of the RNAP till the termination of transcription at the site $i=L$.

Thus, for the special case of the model shown in Fig. \ref{fig-FP_model} the state of the RNAP motor at a particular instant is indicated by the pair $(j,\mu)$, where $j$ is its position on the DNA template i.e $j=1 \, to \, L$ and $\mu$ ($\mu=0, +1, +2, -1$) is the `extra length' of the associated nascent transcript. Fig. \ref{fig-FP_model} clearly shows that, in this special case, a RNAP can follow four different pathways when it encounters a slippery site: (1) the transitions (J-1,0) $\rightarrow$ (J,0) $\rightarrow$ (J+1,0) corresponds to the normal transcription at the slippery site, (2) the transitions (J-1,0) $\rightarrow$ (J,0) $\rightarrow$ (J,+1) $\rightarrow$ (J+1,+1) corresponds to the single backward slippage of transcript at the slippery site, (3) the transitions (J-1,0) $\rightarrow$ (J,0) $\rightarrow$ (J,+1) $\rightarrow$ (J,+2) $\rightarrow$ (J+1,+2) corresponds to the double backward slippage of transcript at the slippery site and (4) the transitions (J-1,0) $\rightarrow$ (J,0) $\rightarrow$ (J,-1) $\rightarrow$ (J+1,-1) corresponds to the single forward slippage of transcript at the slippery site.

\section{Passage of RNAP across slippery site suffering transcript slippage}

\subsection{First Passage Times across slippery site and transient behaviour} 

We define the time $\tau$ taken by the RNAP motor to reach, {\it for the first time}, the position $J+1$,  starting from the position $J-1$ as the {\it first-passage time}. Since the kinetics of transcription, including TS, is probabilistic, $\tau$ varies from one RNAP to another. In this section we calculate the probability distribution (more precisely, the probability density function) $f(\tau)$ of $\tau$. The mean first passage time can be obtained from 
\begin{eqnarray}
\langle \tau \rangle=\int_{0}^{\infty} \tau ~f(\tau)~d\tau
\end{eqnarray}
if $f(\tau)$ is known.

We define $P_{\mu} (J,t)$ as the probability of finding the RNAP in the ``slippage state'' $\mu$ at site $J$ on the DNA track at time $t$. The master equations governing the time evolution of $P_{\mu} (J,t)$ corresponding to the general $N$-state kinetic model , written using the matrix notation, is
\begin{eqnarray}
\frac{d\textbf{P}(t)}{dt}&=&A~ \textbf{P}(t),
\label{matrix-eq1}
\end{eqnarray}
where the vector $\textbf{P}(t)$ is a $N$-component column vector, the components of which are $P_{0} (J-1,t)$, $P_{0} (J+1,t)$ and $P_{\mu} (J,t)$ ($\mu=0, +1, +2, -1$), and the elements of the matrix $A$ are the rates of transitions (more appropriately, the transition probabilities per unit time) between these states. For example, in the special case of the 6-state kinetic model shown in Fig. \ref{fig-FP_model}, defining 
\begin{widetext}
\begin{equation}
\textbf{P}(t) = {\begin{pmatrix}~ P_{0}(J-1,t)~ \\ \\ P_{0}(J,t) \\ \\ P_{0}(J+1,t) \\ \\ P_{+1}(J,t) \\ \\ P_{-1}(J,t) \\ \\~ P_{+2}(J,t)~\end{pmatrix}}
\end{equation}
we have
\begin{eqnarray}
A = 
{\begin{pmatrix}~ 
-q & 0 & 0 & 0 & 0 & 0\\ \\ q & -(q_{0}+b_{1}+f_{1}) & 0 & 0 & 0 & 0\\ \\ 0 & q_{0} & 0 & q_{+1} & q_{-1} & q_{+2} \\ \\ 0 & b_{1}  & 0 & -(q_{+1}+b_{2}) & 0 & 0 \\ \\ 0 & f_{1} & 0 & 0 & -q_{-1} & 0 \\ \\ 0 & 0 & 0 & b_{2} & 0 & -q_{+2}
~\end{pmatrix}}
\label{matrix-eqn} 
\end{eqnarray}
\end{widetext}

Sometimes the computation of the distribution of FPT turns out to be easier 
in terms of Laplace transforms. Carrying out Laplace transform of both sides of Eq. (\ref{matrix-eq1}),
\begin{eqnarray}
\mathcal{L}\biggl[\frac{d\textbf{P}(t)}{dt}\biggr] = A ~\mathcal{L} ~\textbf{P}(t) 
\end{eqnarray}
we get
\begin{eqnarray}
s \tilde{\textbf{P}}(s) - \textbf{P}(0) = A~\tilde{\textbf{P}}(s) 
\end{eqnarray}
and hence
\begin{eqnarray}
\tilde{\textbf{P}}(s)&=&(s \textbf{I} - A) ^{-1}~\textbf{P}(0)
\label{laplace-eq}
\end{eqnarray}
where $\textbf{I}$ is the identity matrix, $\mathcal{L}$ indicates the Laplace transform  operator and $\tilde{\textbf{P}}(s)$ is the Laplace transform of $\textbf{P}(t)$.
In principle, after taking the inverse Laplace transform of $\tilde{\textbf{P}}(s)$, one would get the distribution 
of FPT
\begin{eqnarray}
\textbf{P}(t)&=&\mathcal{L}^{-1}~ [ ~\tilde{\textbf{P}}(s) ~]
\label{eq-invLT}
\end{eqnarray} 
Often the set of kinetic equations is so complicated that the operation of inverse Laplace transform 
(\ref{eq-invLT}) cannot be completed analytically to get a closed-form analytical expression for 
$P_{\mu}(J,t)$. In such situations, the mean first passage time can still be obtained by taking appropriate 
derivatives of $\tilde{P}_{\mu}(J,s)$ if the latter can be calculated in the s-space (Laplace space):
\begin{eqnarray}
\int_{0}^{\infty} t~ P_{\mu} (J,t) ~ dt= - \frac{d}{ds} \tilde{P}_{\mu} (J,s) \Big|_{s=0}
\label{eq-FPT}
\end{eqnarray}

along with the normalization condition  
\begin{eqnarray}
P_{0}(J-1,t)+\sum_{\mu} P_{\mu}(J,t)+P_{0}(J+1,t)=1,
\label{eq-normalization}
\end{eqnarray}

For the calculation of the first-passage time, we impose the initial conditions: 
\begin{equation}
\textbf{P}(0) = {\begin{pmatrix}~ 1~ \\ \\ 0 \\ \\ 0 \\ \\ 0 \\ \\ 0 \\ \\~ 0~\end{pmatrix}}
\label{eq-P0}
\end{equation}
The probability density of first-passage times to reach the target site $J+1$ between time $t$ and $t+dt$ is,
\begin{eqnarray}
f(t)&=&q_{0}~P_{0} (J,t)+q_{+1}~P_{+1} (J,t)+q_{-1}~P_{-1} (J,t) \nonumber\\&&+q_{+2}~P_{+2} (J,t) 
\label{eq-ftexpression}
\end{eqnarray}



\begin{figure}[h]
\begin{center}
(a)\\[0.02cm]
\includegraphics[angle=0,width=0.8\columnwidth]{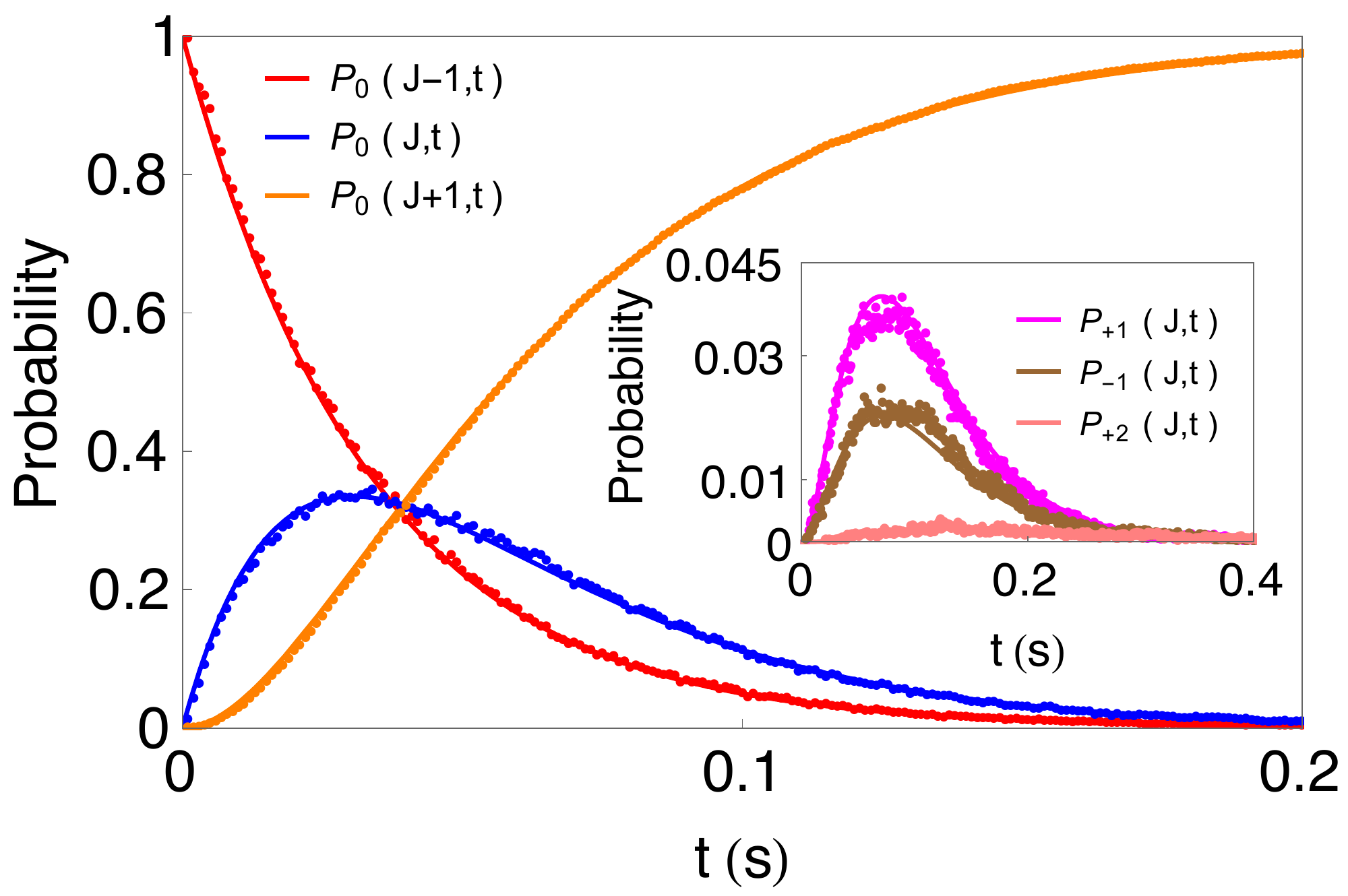}\\[0.02cm]
(b)\\[0.02cm]
\includegraphics[angle=0,width=0.8\columnwidth]{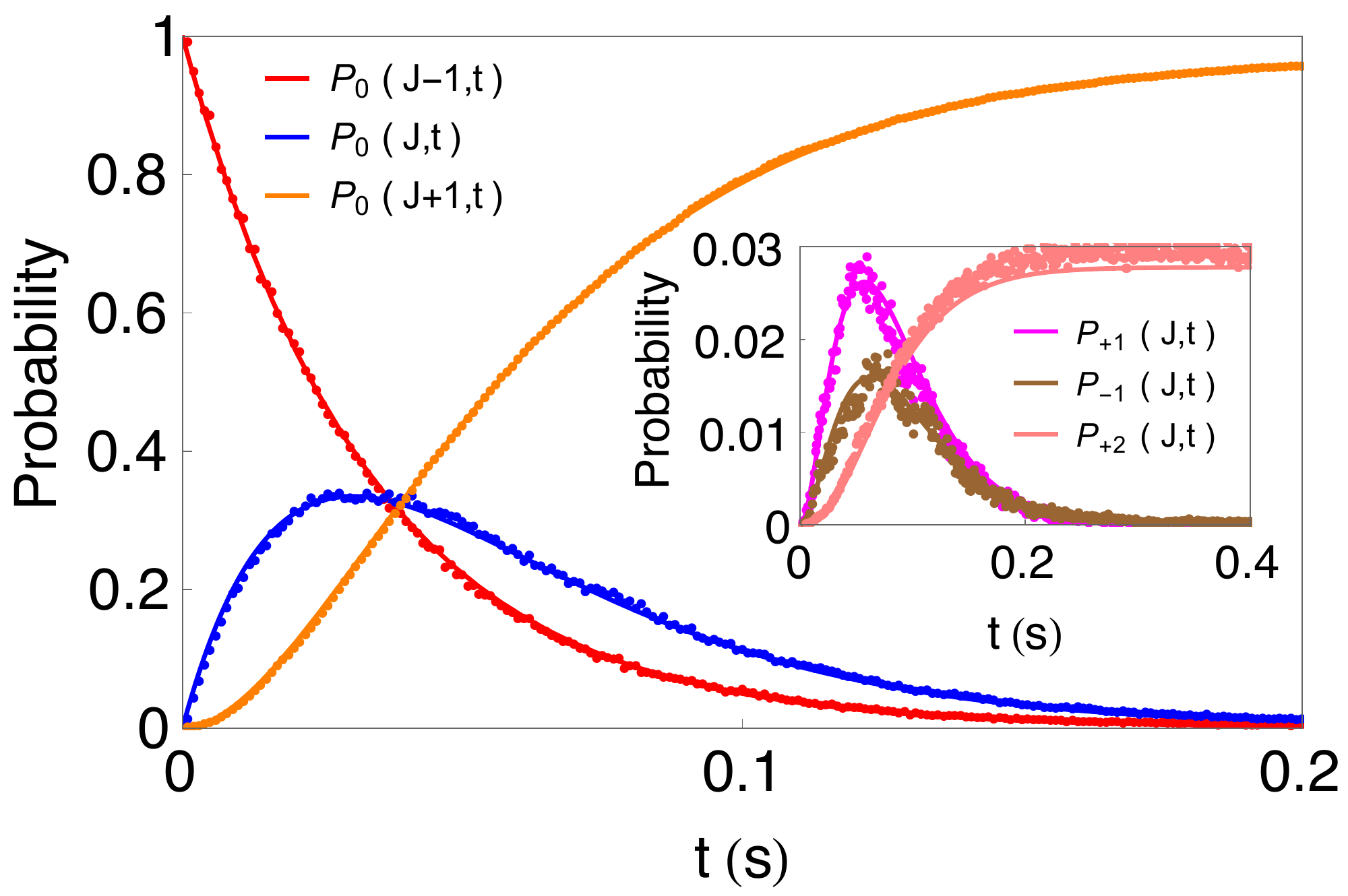}\\[0.02cm]
\end{center}
\caption{
Distribution of probabilities $\textbf{P}(t)$ against time $t$; for (a) $q_{+2}=10~s^{-1}$, $b_{2}=1~s^{-1}$ and (b) $q_{+2}=0~s^{-1}$, $b_{2}=10~s^{-1}$. All the other parameters are kept fixed at values: $q=q_{0}=30~s^{-1}$, $q_{+1}=q_{-1}=20~s^{-1}$, $b_{1}=4~s^{-1}$ and $f_{1}=2~s^{-1}$. Lines correspond to analytical result and discrete data points were obtained from simulation. Inset plot shows how the probabilities of slippage states vary with time.}
\label{fig-P0toP+2}
\end{figure}

The transition matrix $A$ \eqref{matrix-eqn} can be simpliy diagonalized as follows:
\begin{align}
A|\lambda_n\rangle=\lambda_n|\lambda_n\rangle
\end{align}
where the eigenvalues $\lambda_n$ are given by
\begin{align}
&
\lambda_1=0, \quad
\lambda_2=-q, \quad
\lambda_3=-b_1 - f_1 - q_0, 
\nonumber\\&
\lambda_4=-q_{-1}, \quad
\lambda_5=-b_2- q_{+1}, \quad
\lambda_6=-q_{+2}.
\end{align}
The time-dependence of the eigenvectors are simpliy written as
\begin{align}
|\lambda_n(t)\rangle=e^{\lambda_n t}|\lambda_n\rangle. 
\end{align}
Now we expand the initial state $|P(0)\rangle$, given by (\ref{eq-P0}), in terms of the eigenvectors of $A$ as
\begin{align}
|P(0)\rangle=\sum_{n=1}^6 c_n |\lambda_n\rangle, 
\end{align}
where
\begin{align}
&c_1=1, 
\quad
c_2=\frac{b_1 b_2 q}{x_1 x_2 x_3}, 
\quad
c_3=\frac{b_1 b_2 q}{x_1 x_{21} x_{13} }, 
\nonumber\\&
c_4=\frac{f_1 q}{x_4 x_{41}}, 
\quad
c_5=\frac{b_1 b_2 q}{x_2  x_{12} x_{23} }, 
\nonumber\\&
c_6=\frac{b_1 b_2 q}{x_3 x_{13} x_{32}}
\end{align}
and 
\begin{align}
&
x_1=b_1+f_1+q_0-q, \quad
x_2=b_2+q_{+1}-q, 
\nonumber\\&
x_3=q_{+2}-q, \quad
x_4=q_{-1}-q, 
\nonumber\\&
x_{jk}=x_j-x_k, 
\end{align}
from which we have
\begin{align}
|P(t)\rangle=\sum_{n=1}^6 c_n e^{\lambda_n t} |\lambda_n\rangle. 
\end{align}
Explicitly each component can be written as
\begin{align}
&
P_{0}(J-1,t)=e^{-qt}, 
\nonumber\\&
P_{0}(J,t) 
=
qe^{-qt}\left(\frac{1-e^{-x_1t}}{x_1}\right), 
\nonumber\\&
P_{+1}(J,t) 
=
\frac{b_1 q e^{-qt}}{x_{12}}\left(\frac{1-e^{-x_2t}}{x_2}-\frac{1-e^{-x_1t}}{x_1}\right), 
\nonumber\\&
P_{-1}(J,t) 
=
\frac{f_1 q e^{-qt}}{x_{14}}\left(\frac{1-e^{-x_4t}}{x_4}-\frac{1-e^{-x_1t}}{x_1}\right), 
\nonumber\\&
P_{+2}(J,t) 
=
\frac{b_1 b_2 q e^{-qt}}{x_{12}x_{23}x_{13}}
\nonumber\\&\times
\left(
x_{12}\frac{1-e^{-x_3 t}}{x_3}
+x_{23}\frac{1-e^{-x_1 t}}{x_1}
-x_{13}\frac{1-e^{-x_2 t}}{x_2}
\right), 
\label{eq-P0toP+2}
\end{align}
Finally, the expression for the remaining probability $P_{0}(J+1,t)$ can be obtained simply 
using the normalization condition (\ref{eq-normalization}), i.e., 
\begin{align}
P_{0}(J+1,t)
&
=1-\{
P_{0}(J-1,t)+
P_{0}(J,t)
\nonumber\\&
+
P_{+1}(J,t)+
P_{-1}(J,t)+
P_{+2}(J,t) 
\}. 
\label{eq-6thProb}
\end{align}

The exact expressions (\ref{eq-P0toP+2})-(\ref{eq-6thProb}) for the six  probabilities $P_{0}(J-1,t), P_{0}(J,t), P_{0}(J+1,t), P_{+1}(J,t), P_{-1}(J,t)$ and $P_{+2}(J,t)$ are drawn graphically for two sets of the rate constants in 
Fig.\ref{fig-P0toP+2} (a) and (b). On the same graph we also plot the corresponding numerical data obtained from our Monte Carlo simulations of the model with the same set of values of the rate constants. The excellent agreement between the theory and simulation establishes the high accuracy of the simulation data because the analytical expressions (\ref{eq-P0toP+2}) are exact.  The variation of the probabilities with time are consistent with the intuitive expectation based on the initial conditions. The only difference between the Fig.\ref{fig-P0toP+2} (a)  and Fig.\ref{fig-P0toP+2} (b) is that $P_{+2}(J,t)$ saturates to a constant value in the latter whereas it decays to zero in the former. This qualitative difference arises from the choice of parameter value $q_{+2}=0$ in Fig.\ref{fig-P0toP+2} (b) because of which probability $P_{+2}(J,t)$ cannot decay to zero as $t \to \infty$. If $q_{+2}=0$, transcription pauses after two backward slippages. It means that there is no leakage of probability after the second backward slippage, so the corresponding probability cannot decay to zero as $t \to \infty$.

\begin{figure}[h]
\begin{center}
\includegraphics[angle=0,width=0.8\columnwidth]{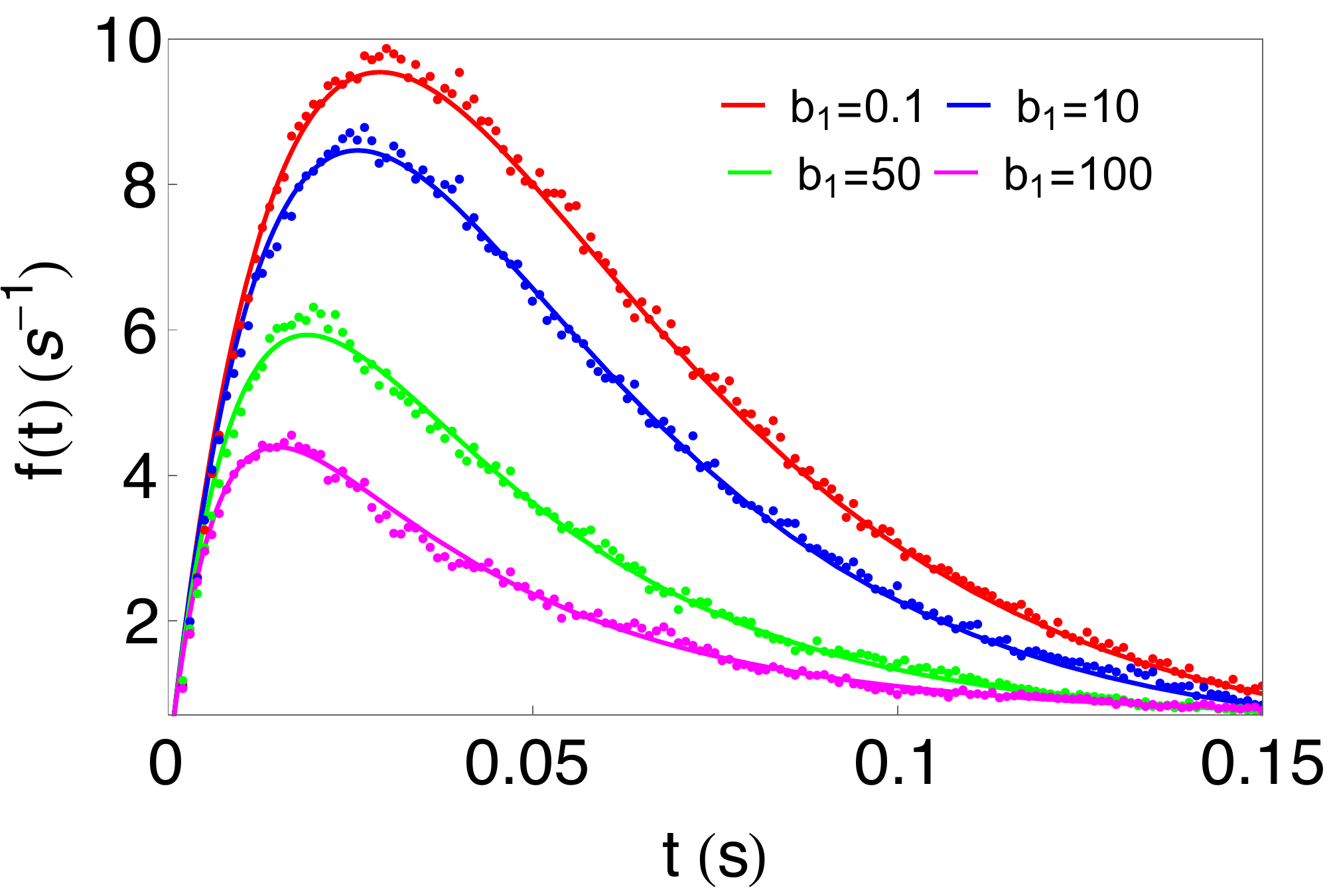}\\[0.02cm]
\end{center}
\caption{Distribution of probability density of first-passage time $f(t)$ as a function of time ($t$); for different values of $b_{1}$. All the other parameters are kept fixed at values: $q=q_{0}=30~s^{-1}$, $q_{+1}=q_{-1}=q_{+2}=1~s^{-1}$, $b_{2}=f_{1}=10~s^{-1}$. Lines correspond to analytical result and discrete data points are obtained from simulation.}
\label{fig-result-probability}
\end{figure}

The exact expression of the distribution of first-passage times can now be 
obtained by substituting (\ref{eq-P0toP+2})-(\ref{eq-6thProb}) into the relation 
(\ref{eq-ftexpression}). Fig. \ref{fig-result-probability} shows the distribution of the {\it first passage time} ($\tau$) for four different values of $b_{1}$  keeping all the other parameter values fixed.

\begin{figure}[h]
\begin{center}
(a)\\[0.02cm]
\includegraphics[angle=0,width=0.8\columnwidth]{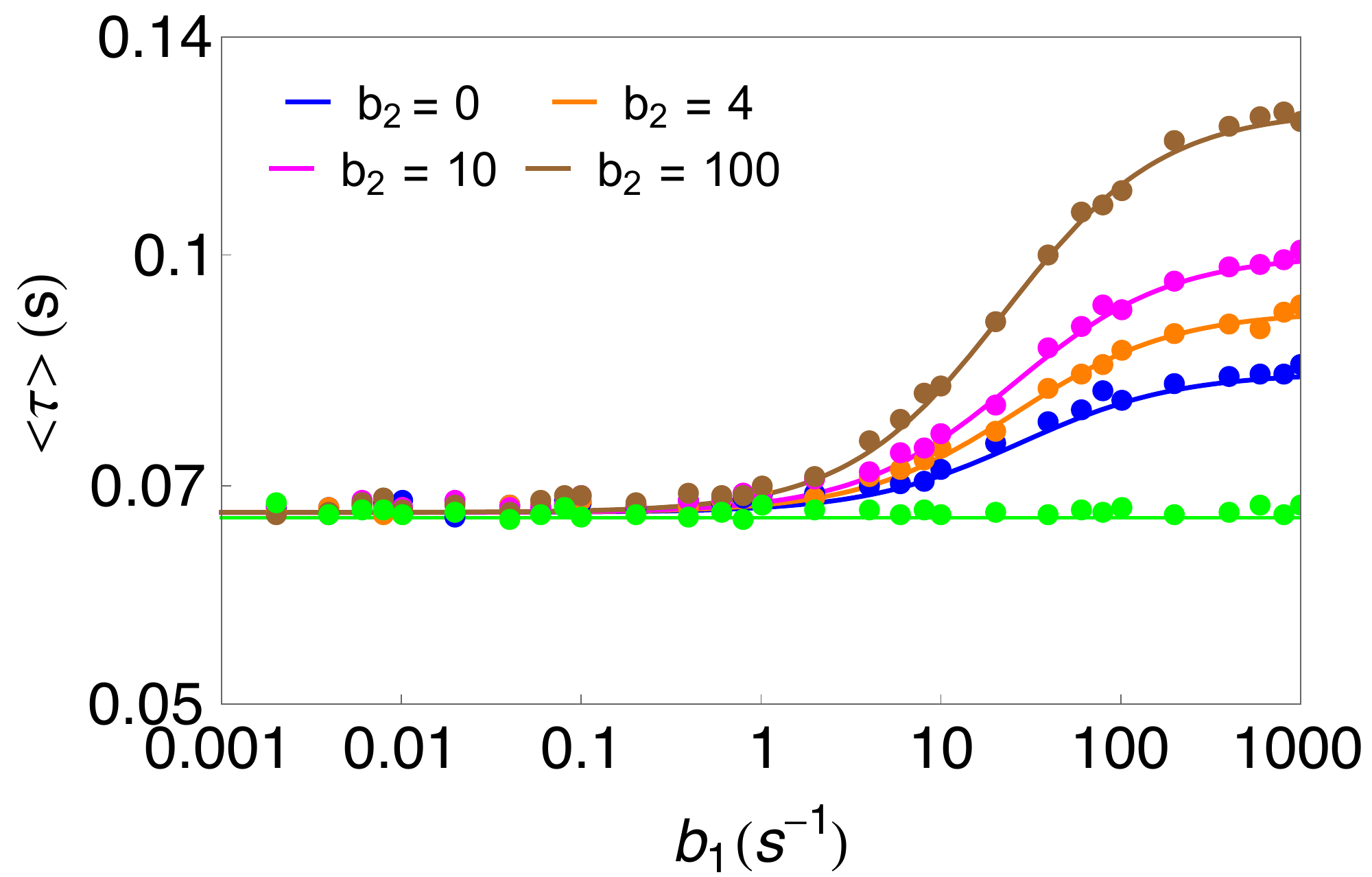}\\[0.02cm]
(b)\\[0.02cm]
\includegraphics[angle=0,width=0.8\columnwidth]{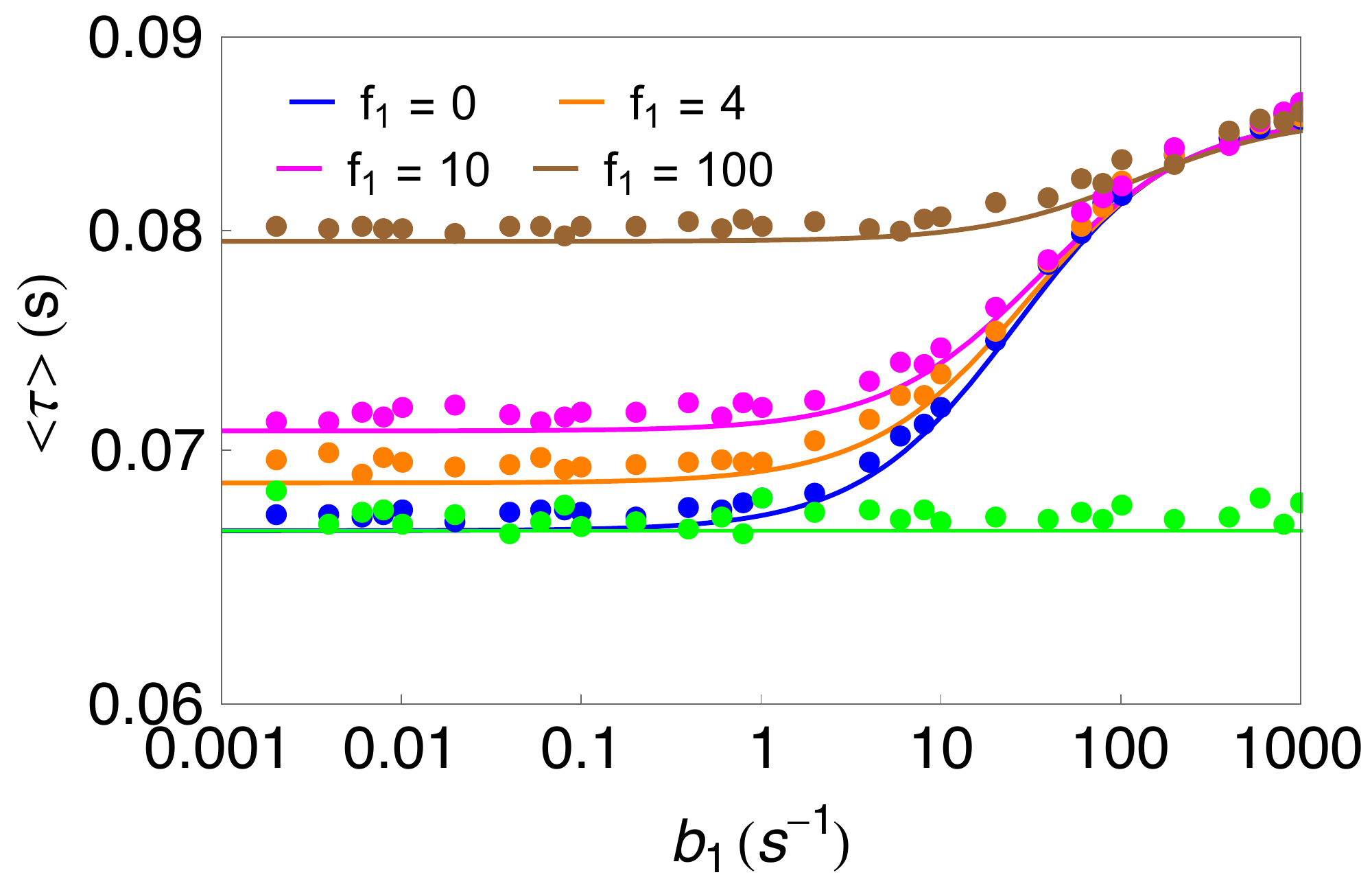}\\[0.02cm]
\end{center}
\caption{Variation of $\langle \tau \rangle$ with respect to variation of first backward slippage rate $b_{1}$; for (a) fixed $f_{1} = 1~s^{-1}$ and (b) fixed $b_{2} = 1~s^{-1}$. All other parameters are kept fixed at values: $q=q_0=30~s^{-1}$, $q_{+1}=q_{-1}=20~s^{-1}$ and $q_{+2}=10~s^{-1}$. Lines correspond to analytical result and points represent the simulation data. Only exception is that green colour line (simulation data) corresponds to $q=q_0=q_{+1}=q_{-1}=q_{+2}=30~s^{-1}$.  }
\label{fig-result_FPT}
\end{figure}

The variation of the mean first-passage time $\langle \tau \rangle$ with $b_{1}$ is shown for four different values of $b_{2}$ in Fig.\ref{fig-result_FPT}(a) and for four different values of $f_{1}$ in Fig.\ref{fig-result_FPT}(b). The higher is the rate of slippage the longer it takes for the RNAP to pass the defect site.

\subsection{Steady state: fractions of slipped transcripts} 

Since no TS is assumed to occur at the $L-1$ sites labelled by $j \neq J$, the lengths of the transcripts 
synthesized by the RNAPs in the 6-state model can have only the lengths $L$, $L-1$, $L+1$ and $L+2$.  
For the steady state of the system, we can define the corresponding probabilities by the relations
\begin{eqnarray}
P_{L}&=&\frac{P_{0}(J+1)}{P(J+1)}=\frac{q_{0}}{b_{1} + f_{1} + q_{0}} \nonumber \\
P_{L+1}&=&\frac{P_{+1}(J+1)}{P(J+1)}=\frac{b_{1}q_{+1}}{(b_{1} + f_{1} + q_{0})(b_{2}+q_{+1})} \nonumber \\
P_{L-1}&=&\frac{P_{-1}(J+1)}{P(J+1)}=\frac{f_{1}}{b_{1} + f_{1} + q_{0}} \nonumber \\
P_{L+2}&=&\frac{P_{+2}(J+1)}{P(J+1)}=\frac{b_{1}b_{2}}{(b_{1} + f_{1} + q_{0})(b_{2}+q_{+1})} \nonumber \\
\label{eq-trlength}
\end{eqnarray}
where 
\begin{equation}
P(J+1) = P_{0}(J+1) + P_{+1}(J+1) + P_{+2}(J+1) + P_{-1}(J+1).
\end{equation}
Thus, $P_{L}, P_{L+1}, P_{L-1}, P_{L+2}$ can be interpreted as the fractions of the respective species 
of the transcripts synthesized. The expressions (\ref{eq-trlength}) for the probabilities depend only on the position, but not on time $t$, because these correspond to the NESS which is attained in the limit $t \to \infty$.


\begin{figure}[h]
\begin{center}
\includegraphics[angle=0,width=0.9\columnwidth]{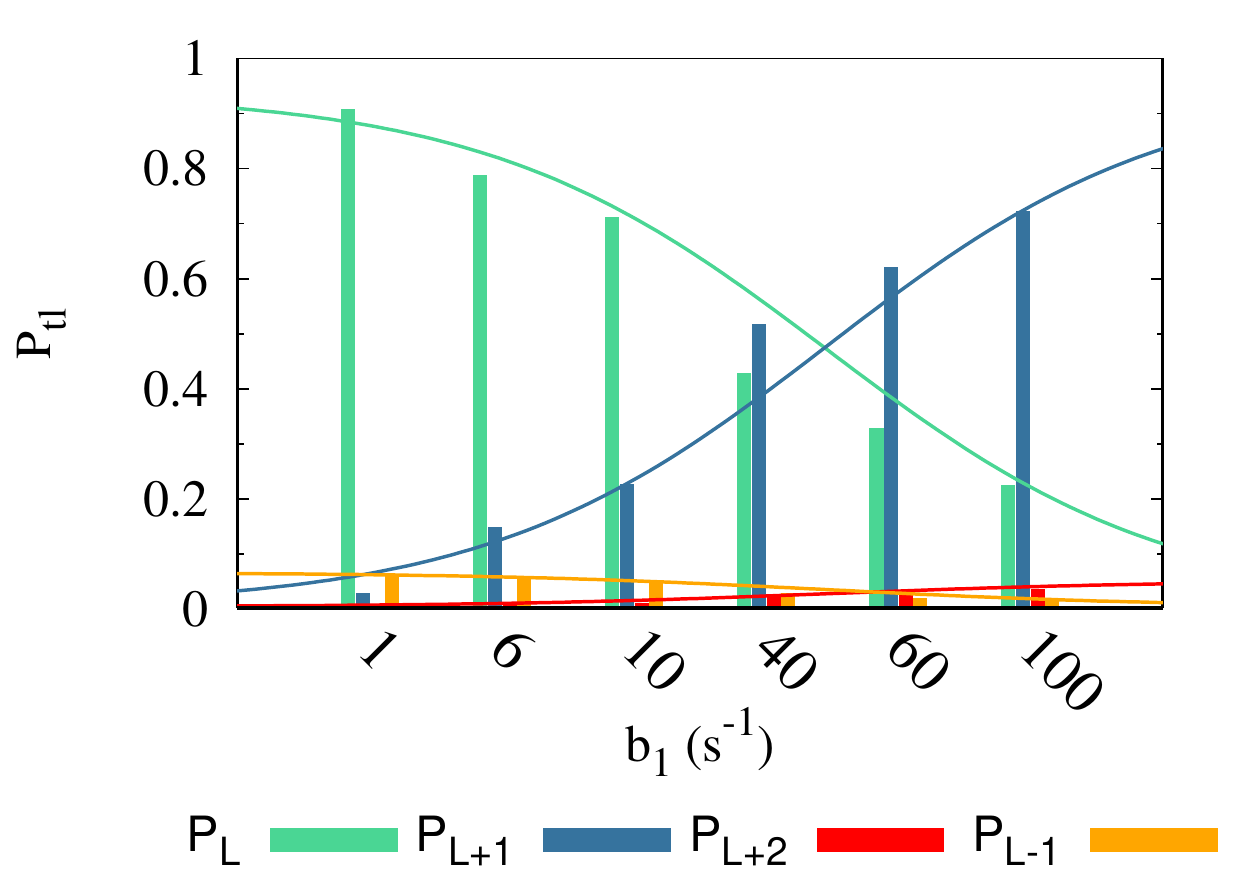}\\[0.02cm]
\end{center}
\caption{Histogram of probability of types of nascent mRNA are plotted with different values of first backward slippage rate $b_{1}$. Parameter values are $b_{2} = 1~s^{-1}$ and $f_{1} = 2~s^{-1}$. All other parameters are kept fixed at values:  $q=q_0=30~s^{-1}$, $q_{+1}=q_{-1}=20~s^{-1}$ and $q_{+2}=10~s^{-1}$. Lines correspond to exact expressions (\ref{eq-trlength}), where green, blue, red and orange continuous lines correspond to $P_{L}$, $P_{L+1}$, $P_{L+2}$ and $P_{L-1}$, respectively. Bar plots are obtained from simulation data. }
\label{fig-result_histogram}
\end{figure}

Fig. \ref{fig-result_histogram} shows the histogram for probability of transcript lengths ($P_{tl}$) plotted against single backward slippage rate, $b_{1}$. For low values of $b_{1}$, $P_{L}$ is higher than $P_{L+1}$, $P_{L+2}$ and $P_{L-1}$, as the chances of backward slippages are lower than that of forward slippage. For high values of $b_{1}$, possibility of backward slippage gets enhanced and hence $P_{L+1}$ is higher than $P_{L+2}$, $P_{L-1}$ and $P_{L}$. Our analytical results show that the probability of $P_{L+1}$ and $P_{L+2}$ increases as $b_{1}$ increases and $P_{L}$ and $P_{L-1}$ decreases as $b_{1}$ increases. The trend in the histogram plot qualitatively matches with recent experimental findings \cite{olspert15,olspert16}. \\

\section{Effect of RNAP traffic congestion on the transcript slippage phenomenon}

\begin{figure}[h]
\begin{center}
\includegraphics[angle=0,width=0.95\columnwidth]{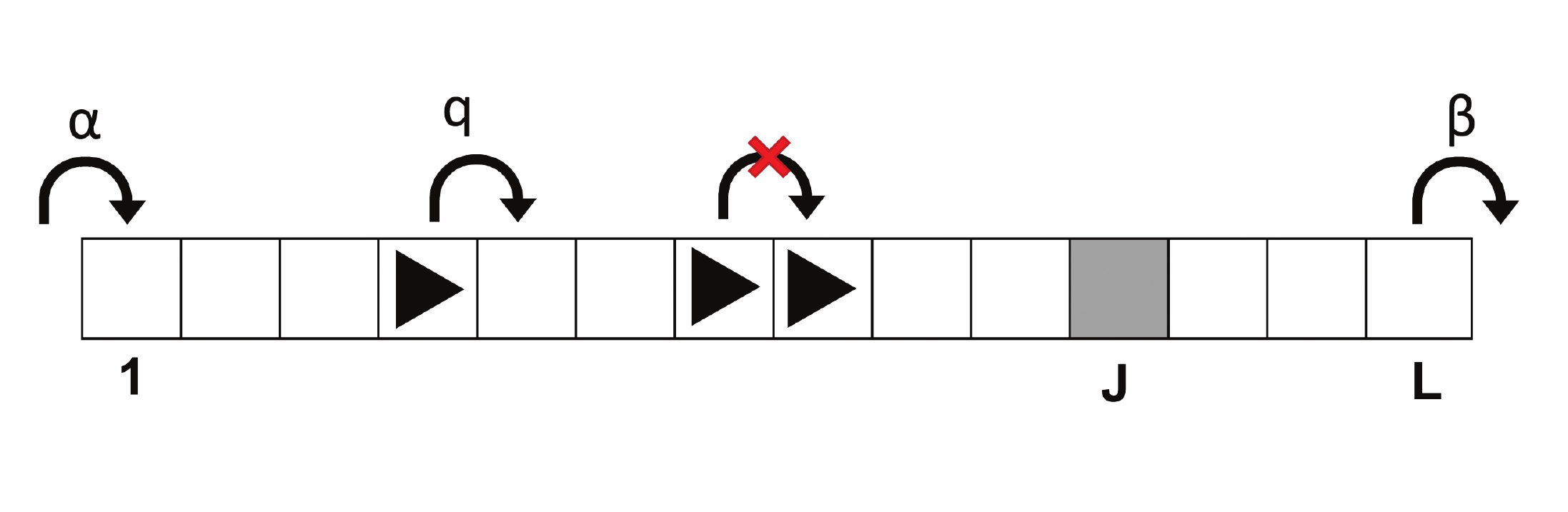}\\[0.02cm]
\end{center}
\caption{Schematic diagram of RNAP traffic on the DNA track of length L in presence of slippage at site J. Triangles
represent (single walking) RNAPs. RNAPs can enter the DNA track only at site $i=1$ with rate $\alpha$ if the entry site is empty and RNAPs can leave the DNA track after it reaches the termination site $i=L$ with rate $\beta$. In between, RNAPs can hop forward, if the target site is empty, with the normal transcription rate $q$ except at site J, where it can hop with rates $q_\mu~ (\mu=0,\pm 1,+2)$, depending on the slippage state (refer to Fig. \ref{fig-FP_model})}
\label{fig-meantime_model}
\end{figure}

Fig. \ref{fig-meantime_model} shows a schematic diagram of the model for RNAP traffic on the DNA track in the presence of a slippery site at $J$. A RNAP can enter the DNA track, with rate $\alpha$, if the entry site $i=1$ is empty. If the RNAP is located at any other position $i\neq J$, it can move forward, with rate $q$ if, and only if, the target site is empty. On the other hand while located at the special site $J$ a RNAP can hop forward with rates $q_\mu~ (\mu=0,\pm 1, +2)$, depending on the slippage state (refer to Fig. \ref{fig-FP_model}). A RNAP can detach from the track at the exit site $i=L$ with rate $\beta$. At the slippery site $J$, the nascent RNA can slip backward with rates $b_1$, $b_2$, etc. and forward with the rate $f_{1}$. 
Since TS does not involve any movement of the RNAP with respect to its DNA track, forward slippage (of the RNA transcript) can happen even when the next site in front of the RNAP is covered by another RNAP.

Let $P_{\mu}(i,t)$ denote the probability of finding RNAP in slippage state $\mu$ at site $i$ on the DNA track at time $t$. So, the probability that the site $i$ is occupied by a RNAP at time $t$, irrespective of its slippage state, is $P(i,t)=\sum_{\mu}P_{\mu}(i,t)$, where $\mu=0, +1, +2, -1$. We can refer to this model as a biologically motivated extension of the TASEP with a special kind of defect located at the specific site $i=J=L/2$. Under mean field approximation, the master equation for the probabilities $P_{\mu}(i,t)$ are given by
\begin{widetext}
\begin{eqnarray}
\frac{dP(1,t)}{dt}&=& \alpha(1-P(1,t)) - q P(1,t) (1-P(2,t))\nonumber\\
\frac{dP(i,t)}{dt}&=& q[P(i-1,t)(1-P(i,t)) - P(i,t) (1-P(i+1,t))]~~~{\rm for} ~ 1 < i < L~ ( i \neq L/2, L/2+1 )\nonumber\\
\frac{dP_{0}(L/2,t)}{dt}&=&q P(L/2-1,t) (1-P(L/2,t))-q_0 P_{0}(L/2,t) (1-P(L/2+1,t))-(b_{1}+f_{1}) P_{0}(L/2,t)\nonumber\\
\frac{dP_{+1}(L/2,t)}{dt}&=&b_{1} P_{0}(L/2,t)-q_{+1} P_{+1}(L/2,t) (1-P(L/2+1,t))-b_{2} P_{+1}(L/2,t)\nonumber\\
\frac{dP_{+2}(L/2,t)}{dt}&=&b_{2} P_{+1}(L/2,t)-q_{+2} P_{+2}(L/2,t) (1-P(L/2+1,t))\nonumber\\
\frac{dP_{-1}(L/2,t)}{dt}&=&f_{1} P_{0}(L/2,t)-q_{-1} P_{-1}(L/2,t) (1-P(L/2+1,t))\nonumber\\
\frac{dP(L/2+1,t)}{dt}&=&[q_0 P_{0}(L/2,t)+q_{+1} P_{+1}(L/2,t)+q_{+2} P_{+2}(L/2,t)+q_{-1} P_{-1}(L/2,t)] (1-P(L/2+1,t)) \nonumber\\
& & -q P(L/2+1,t) (1-P(L/2+2,t))\nonumber\\
\frac{dP(L,t)}{dt}&=& q P(L-1,t) (1-P(L,t)) - \beta P(L,t).\nonumber\\
\end{eqnarray}
\end{widetext}

In the steady state the left hand sides of all these equations vanish and the corresponding solutions of the equations are obtained iteratively by checking whether the difference of the numerical values of two successive iterations is less than $\epsilon$ where $\epsilon \approx 10^{-8}$ is a preassigned small number. 

In our  Monte Carlo simulation of the model, starting from an initial condition, the flux was monitored in each run to ensure that the system reaches a steady state where the flux becomes independent of time. Then, starting from an even longer instant of time $t_{steady}$ ($\equiv n_{steady}~ dt$) the numerical data from the simulation were recorded for the computation of the steady-state properties; the collection of the data were continued until a time $t_{max}$ ($\equiv n_{max}~ dt$)  where the simulation run was terminated. The symbols $n_{steady}$ and $n_{max}$ refer to the corresponding number of Monte Carlo steps and each Monte Carlo step corresponds to the infinitesimal real time interval $dt$. 
The choice of the precise value of $dt$ is based on the fastest rate in the model as the corresponding probability of occurrence of the event must always be less than one. In our Monte Carlo simulations we chose the numerical value  $dt = 5 \times 10^{-4}~s$. 

For the correspondence between the results obtained for arbitrary rates $\alpha$ and $\beta$ (with dimensions of inverse time) and the well known standard results for the TASEP in terms of dimensionless probabilities, where the hopping probability is taken to be unity, requires dividing the rates $\alpha$ and $\beta$ by the rate $q$. In other words, the correspondence requires  $\alpha \rightarrow \alpha'=\alpha/q$ and $\beta \rightarrow \beta'=\beta/q$.
In low density (LD) phase, density is determined by $\alpha'$ ($\rho=\alpha'$) and hence $\alpha'<1/2<\beta'$. In  high density (HD) phase, density is determined by $\beta'$ $(\rho=1-\beta'$) and hence $\alpha'>1/2>\beta'$. In maximal current (MC) phase, $\rho=1/2$ and hence $\alpha'=\beta' > 1/2$.

\subsection{Effects of traffic congestion on extent of TS} 

\begin{figure}[h]
\begin{center}
\includegraphics[angle=0,width=0.9\columnwidth]{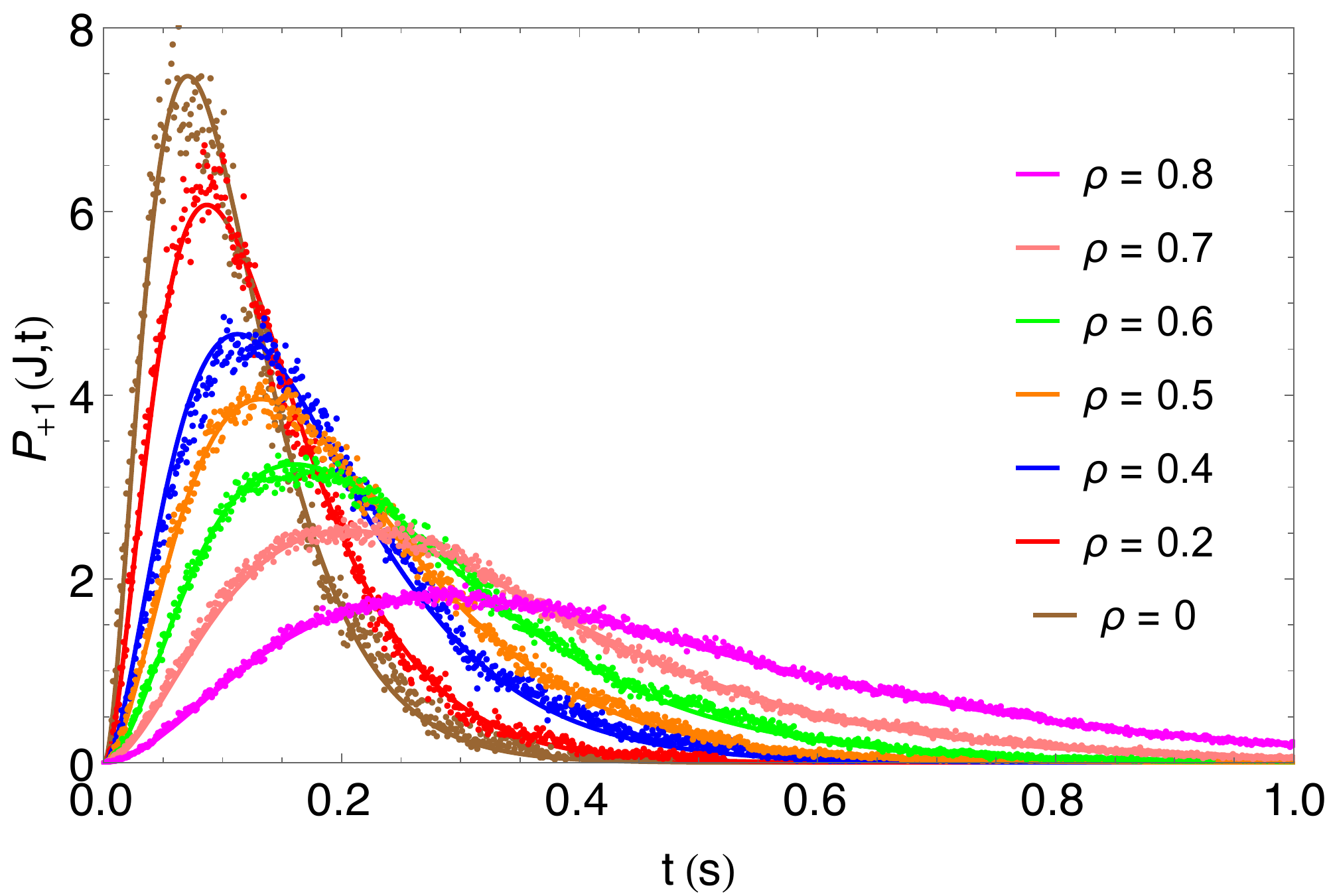}\\[0.02cm]
\end{center}
\caption{Distributions of probability of first backward slippage state, $P_{+1}(J,t)$ with time $t$, for different values of $\rho$. Parameter values for slippage rates: $b_{1}=4~s^{-1}$, $b_{2}=1~s^{-1}$ and $f_{1}=2~s^{-1}$. All other parameters were kept fixed at values: $q=q_0=30~s^{-1}$, $q_{+1}=q_{-1}=20~s^{-1}$ and $q_{+2}=10~s^{-1}$. Lines correspond to analytical result and discrete data points were obtained from simulation.}
\label{fig-result_forward_slippage_state}
\end{figure}

In this subsection, we compute the  mean time taken by each RNAP to transcribe a DNA template of length $L$, on which the defect (i.e., the slippery site) is located at $J$, in the steady state of the RNAP traffic. The simplest way to account for the traffic congestion is to replace the hopping rates of RNAPs,  from one site to the next, by effective rates obtained by multiplying the actual rate with the factor $(1-\rho)$, where $\rho$ is the number density of the RNAPs.

Fig. \ref{fig-result_forward_slippage_state} shows the variation in probability distributions of first backward slippage state plotted against time for different values of $\rho$. The trend observed in the graph is due to crowding i.e, the RNAPs have to face hindrance to move forward.

\begin{figure}[h]
\begin{center}
(a)\\[0.02cm]
\includegraphics[angle=0,width=0.9\columnwidth]{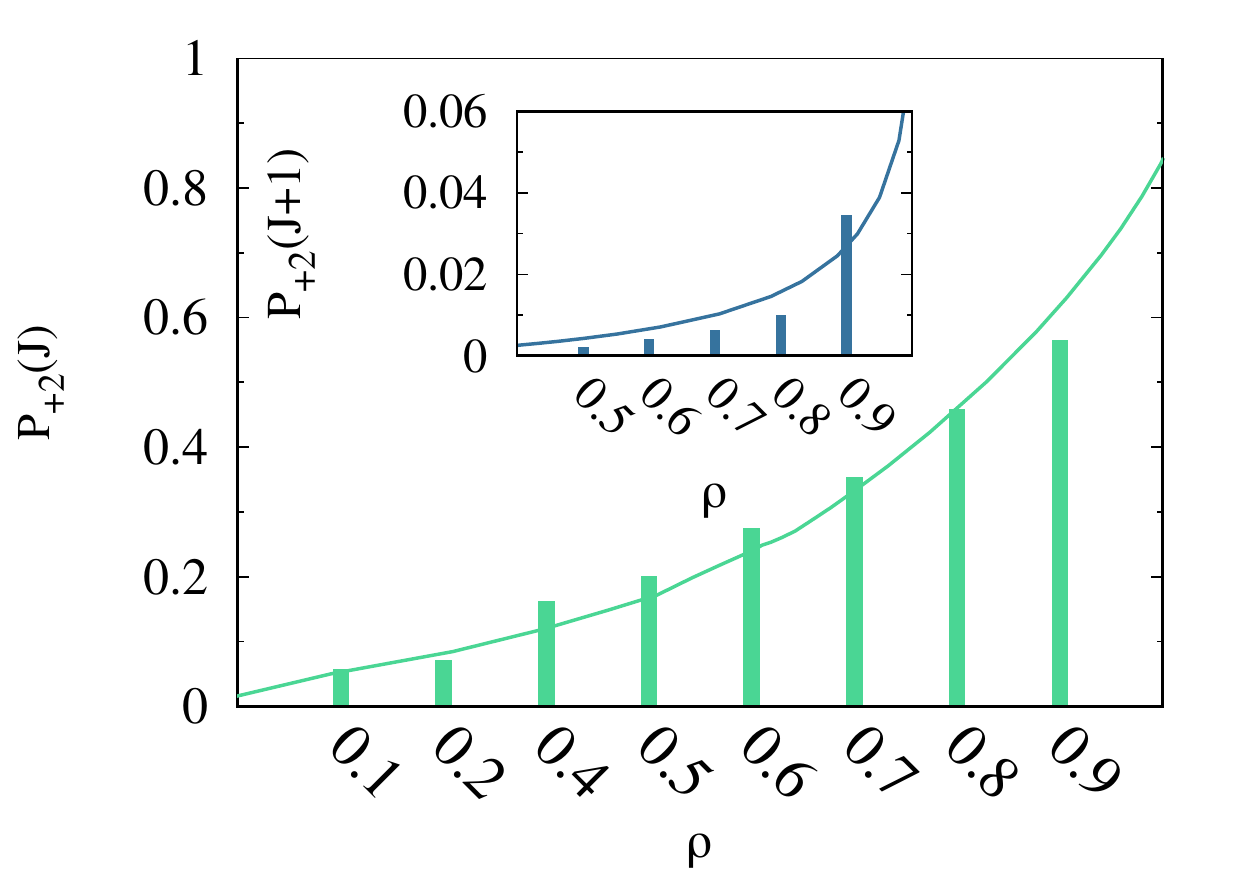}\\[0.02cm]
\end{center}
\caption{The steady state probabilities $P_{+2}(J)$ and $P_{+2}(J+1)$ are plotted in the main figure and inset, respectively, as functions of the number density $\rho$. The parameter values chosen for this figure are 
$b_1=f_1=q_{+2}=0.3$s$^{-1}$ and $q=q_0=q_{+1}=q_{-1}=b_2=30$s$^{-1}$. Lines correspond to MF result and Bar plots were obtained from simulation data.}
\label{fig-enhancedTSrho}
\end{figure}

The denser is the traffic congestion, the longer is the dwell time of an arbitrary RNAP at the slippery site 
and the larger is the expected number of TS events that can occur during the duration of that dwell. 
This intuitive expectation is, indeed, supported by the data shown in Fig.\ref{fig-enhancedTSrho} where 
 $P_{+2}(J)$ and $P_{+2}(J+1)$ have been plotted as functions of the number density $\rho$. 

For plotting this figure we have chosen $b_1=f_1=q_{+2}=0.3$s$^{-1}$ and $q=q_0=q_1=q_{-1}=b_2=30$s$^{-1}$. Because of the 
small values of $b_{1}$ and $f_{1}$, the likelihood of the first TS event, irrespective of forward or 
backward, is normally quite low. However, as the value of $\rho$ increases, the dwell times of 
the RNAPs increase at all sites, including that located at the slippery site. Consequently, during 
that longer period of stay at the slippery site, the RNAP  suffers multiple rounds of TS; this is 
reflected in the increase in the magnitude of $P_{+2}(J)$ in Fig.\ref{fig-enhancedTSrho}. 

\begin{figure}[h]
\begin{center}
(a)\\[0.02cm] 
\includegraphics[angle=0,width=0.9\columnwidth]{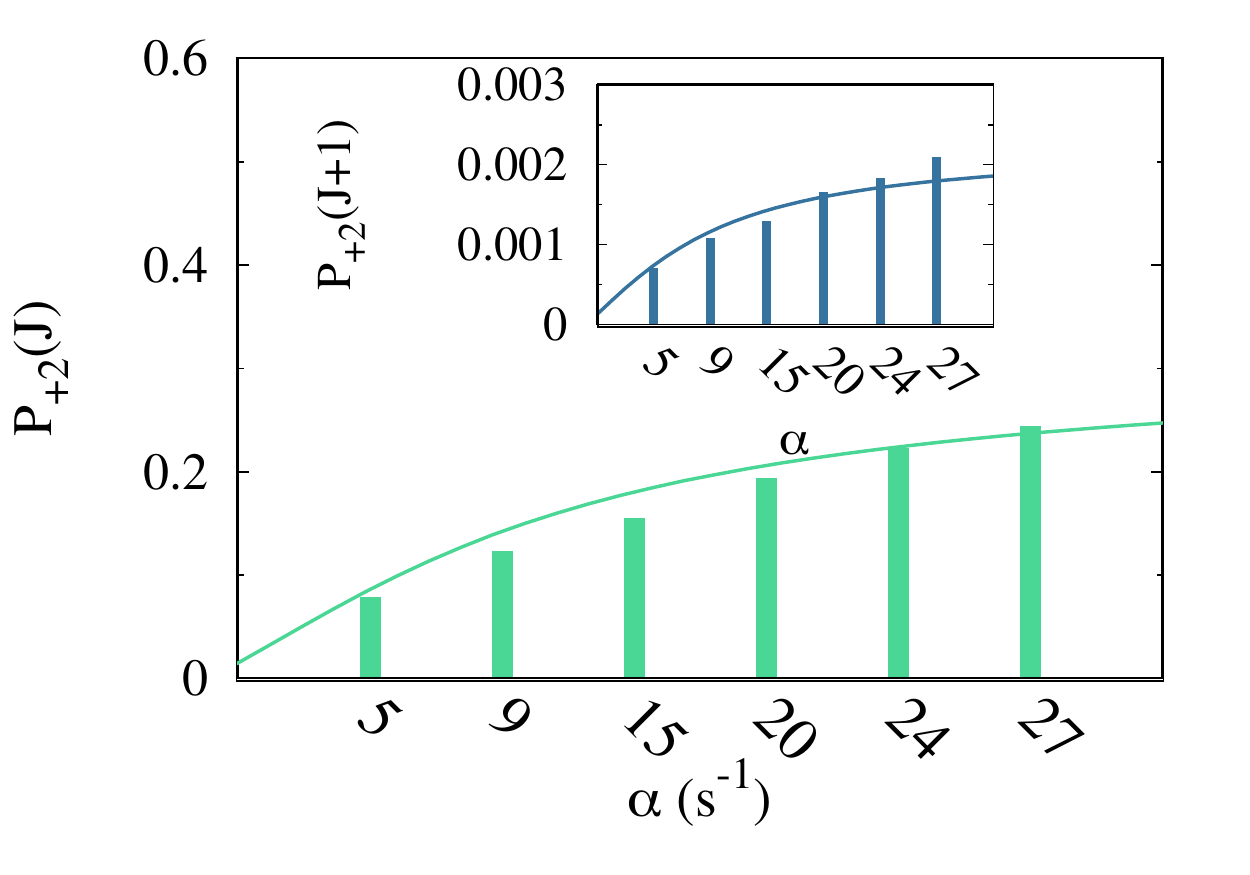}\\[0.02cm]
(b)\\[0.02cm] 
\includegraphics[angle=0,width=0.9\columnwidth]{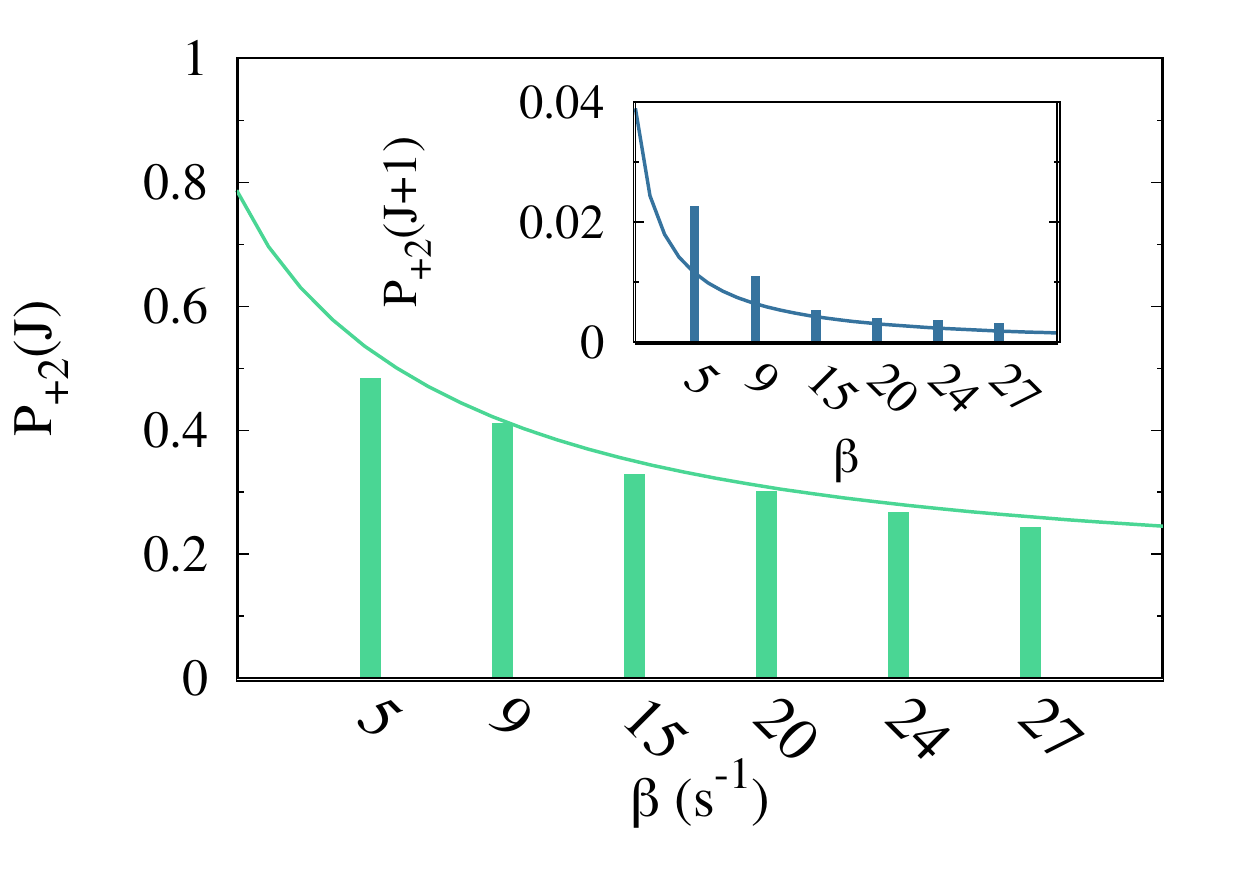}\\[0.02cm]
\end{center}
\caption{The steady state probabilities $P_{+2}(J)$ and $P_{+2}(J+1)$ are plotted in the main figure and inset, respectively, as functions of entry rate $\alpha$ of RNAP (in (a)) and  exit rate $\beta$ of RNAP (in (b)). For both the figures the parameter values chosen  are $b_1=f_1=q_{+2}=0.3$s$^{-1}$ and $q=q_0=q_{+1}=q_{-1}=b_2=30$s$^{-1}$; $\beta=30 s^{-1}$ in (a) and 
$\alpha=30 s^{-1}$ in (b). Lines correspond to MF result and Bar plots were obtained from simulation data.}
\label{fig-enhancedTSalphabeta}
\end{figure}

The probability $P_{+2}(J)$ and P$_{+2}(J+1)$ are plotted as functions of $\alpha$ (for fixed $\beta$) in Fig.\ref{fig-enhancedTSalphabeta}(a) and as as a function of $\beta$ (for fixed $\alpha$) in 
Fig.\ref{fig-enhancedTSalphabeta}(b). As $\alpha$ increases both $P_{+2}(J)$ and P$_{+2}(J+1)$ increase but the  rate of increase decreases gradually and, probabilities eventually saturate because the RNAP traffic makes a transition from the LD phase to the MC phase where the flux of RNAPs saturates. Similarly, for a fixed $\alpha$, as $\beta$ increases the transition from the HD phase to MC reduces the effective dwell time of each RNAP at the defect site which, in turn, reduces the probabilities of multiple TS events at that site. Moreover, the transition to the MC phase also leads to the saturations of the values of $P_{+2}(J)$ and P$_{+2}(J+1)$ with increasing $\beta$.

\begin{figure}[h]
\begin{center}
\includegraphics[angle=0,width=0.9\columnwidth]{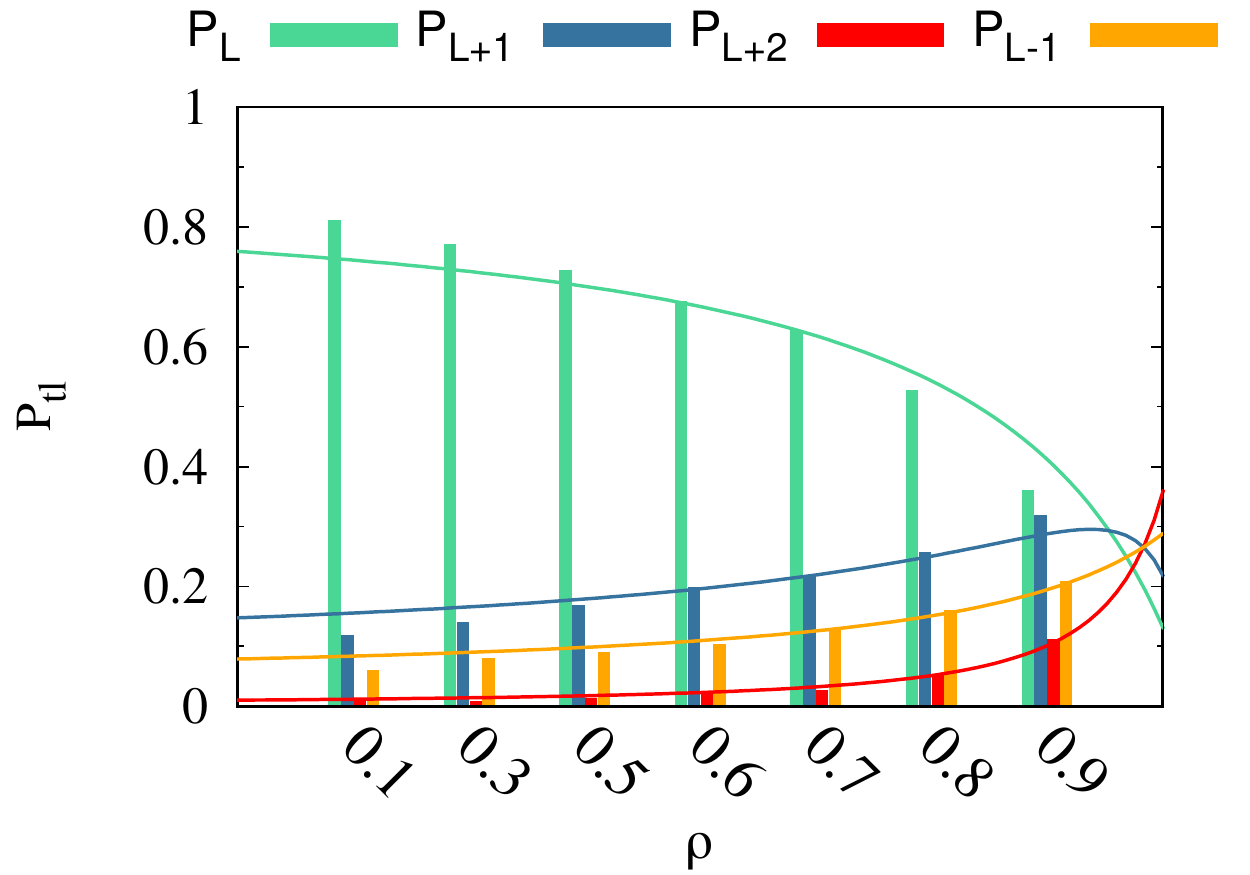}\\[0.02cm]
\end{center}
\caption{Histogram of slippage statistics are plotted with different values of RNAPs density ($\rho$). Parameter values for slippage rates: $b_{1} = 4~s^{-1}$, $b_{2} = 1~s^{-1}$ and $f_{1} = 2~s^{-1}$. All other parameters are kept fixed at values: $q=q_0=30~s^{-1}$, $q_{+1}=q_{-1}=20~s^{-1}$ and $q_{+2}=10~s^{-1}$. Lines correspond to {\it MF} result, where green, blue, red and orange continuous lines correspond to $P_{L}$, $P_{L+1}$, $P_{L+2}$ and $P_{L-1}$, respectively. Bar plots were obtained from simulation data. }
\label{fig-result_histogram_slippage}
\end{figure}

Fig. \ref{fig-result_histogram_slippage} shows the histogram for slippage statistics plotted against $\rho$. In MC phase $(\rho=0.5)$ and even in the LD phase $(\rho<0.5)$, $P_{L+1}$ and $P_{L-1}$ have significantly low values and remain unaffected by the change in $\rho$. In HD phase $(\rho>0.5)$, $P_{L}$ decreases and $P_{L+1}$, $P_{L+2}$ and $P_{L-1}$ increases due to the crowding effect. Because of hindrance, RNAPs have to wait longer time at the slippage site and it enhances the chance of slippages.

\section{Effect of transcript slippage on RNAP traffic flow} 

The time taken by a RNAP, on the average, in the steady state of the RNAP traffic to transcribe can be written as the inverse of the exit rate $\beta$ multiplied by $P_{0}(L)+P_{+1}(L)+P_{+2}(L)+P_{-1}(L)$. So,
\begin{eqnarray}
T_{ss}&=& \frac{1}{\beta P(L)} \nonumber\\
&=& \frac{1}{\beta  (P_{0}(L)+P_{+1}(L)+P_{+2}(L)+P_{-1}(L) )} \nonumber\\
\label{eq-Tss}
\end{eqnarray} 
We obtained the mean-field theoretic estimate of $T_{ss}$ by substituting the mean-field values of the probabilities in the denominator of (\ref{eq-Tss}).

For the computation of $T_{ss}$ in our  Monte Carlo simulation, we use the formula
\begin{eqnarray}
 T_{ss} &=&  \frac{(n_{max} - n_{steady}) dt}{N}
\end{eqnarray}
where, $N$ is the total number of departing RNAPs counted at $i=L$ over the (update) step numbers $n_{max} - n_{steady}$ of the simulation and $dt$ is the duration of each time step of the simulation. We have taken $\alpha=\beta=30~s^{-1}$ (Maximal current phase) and $dt=5 \times 10^{-4}~s=$ time step for  Monte Carlo simulation and MF approximation. We have taken the length of the DNA track (L) to be of 1000 lattice sites.

\begin{figure}[h]
\begin{center}
(a)\\[0.02cm]
\includegraphics[angle=0,width=0.9\columnwidth]{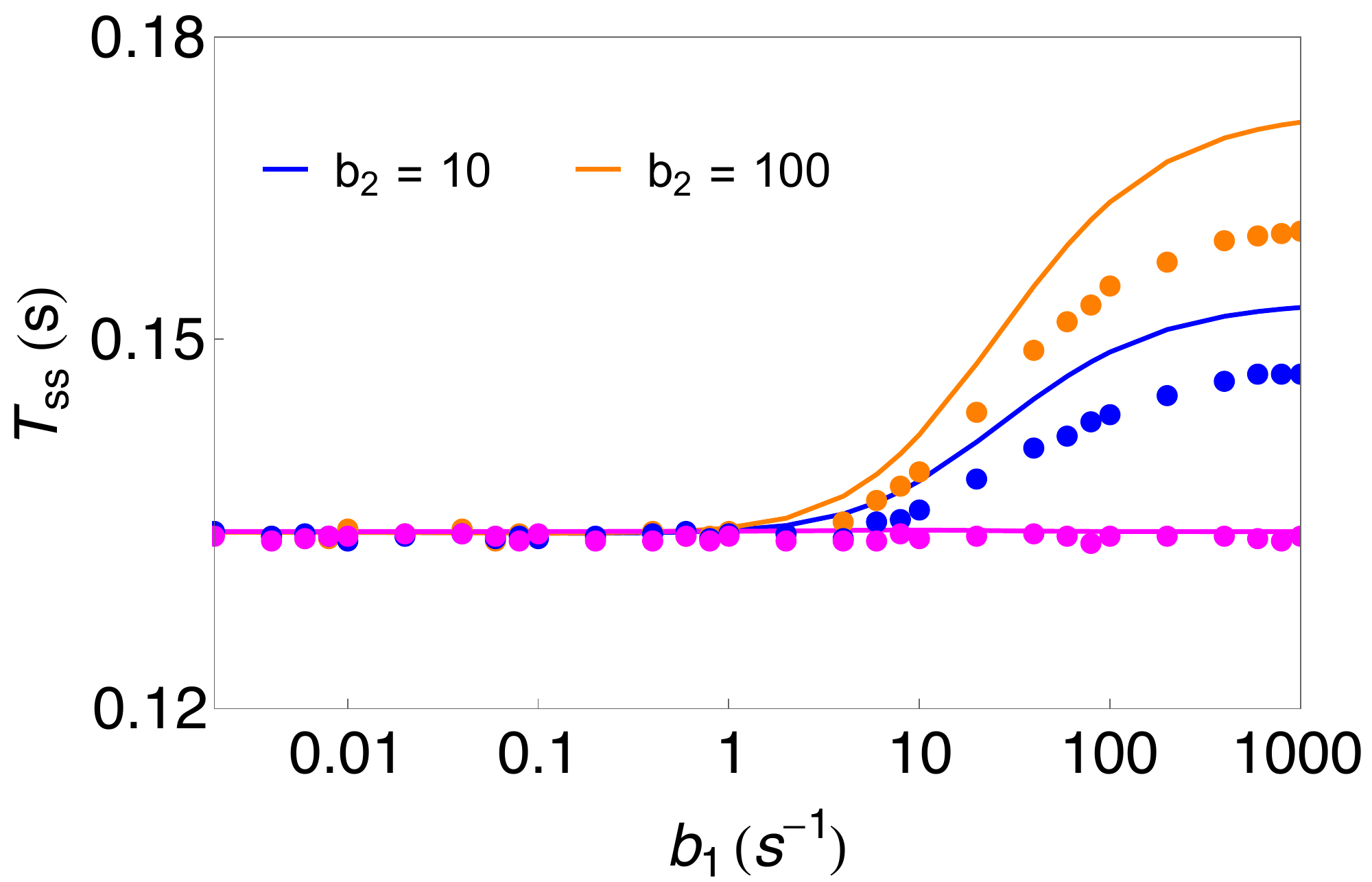}\\[0.02cm]
\end{center}
\caption{Variation of the mean transcription time in steady state ($T_{ss}$) with $b_{1}$ keeping all the other parameters fixed at values  $q=q_0=30~s^{-1}$, $q_{+1}=q_{-1}=20~s^{-1}$, $q_{+2}=10~s^{-1}$ and  $f_{1}=1~s^{-1}$; $\alpha=\beta=30~s^{-1}$ (Maximal current phase) and $dt=5 \times 10^{-4}~s=$ time step for Monte Carlo simulation and MF approximation. Lines correspond to MF theory and points have been obtained from  Monte Carlo simulations. Only exception is that the results shown in magenta colour corresponds to parameter values $q=q_0=q_{+1}=q_{-1}=q_{+2}=30~s^{-1}$. }
\label{fig-result_meantime}
\end{figure}

\begin{figure}[h]
\begin{center}
(a)\\[0.02cm]
\includegraphics[angle=0,width=0.9\columnwidth]{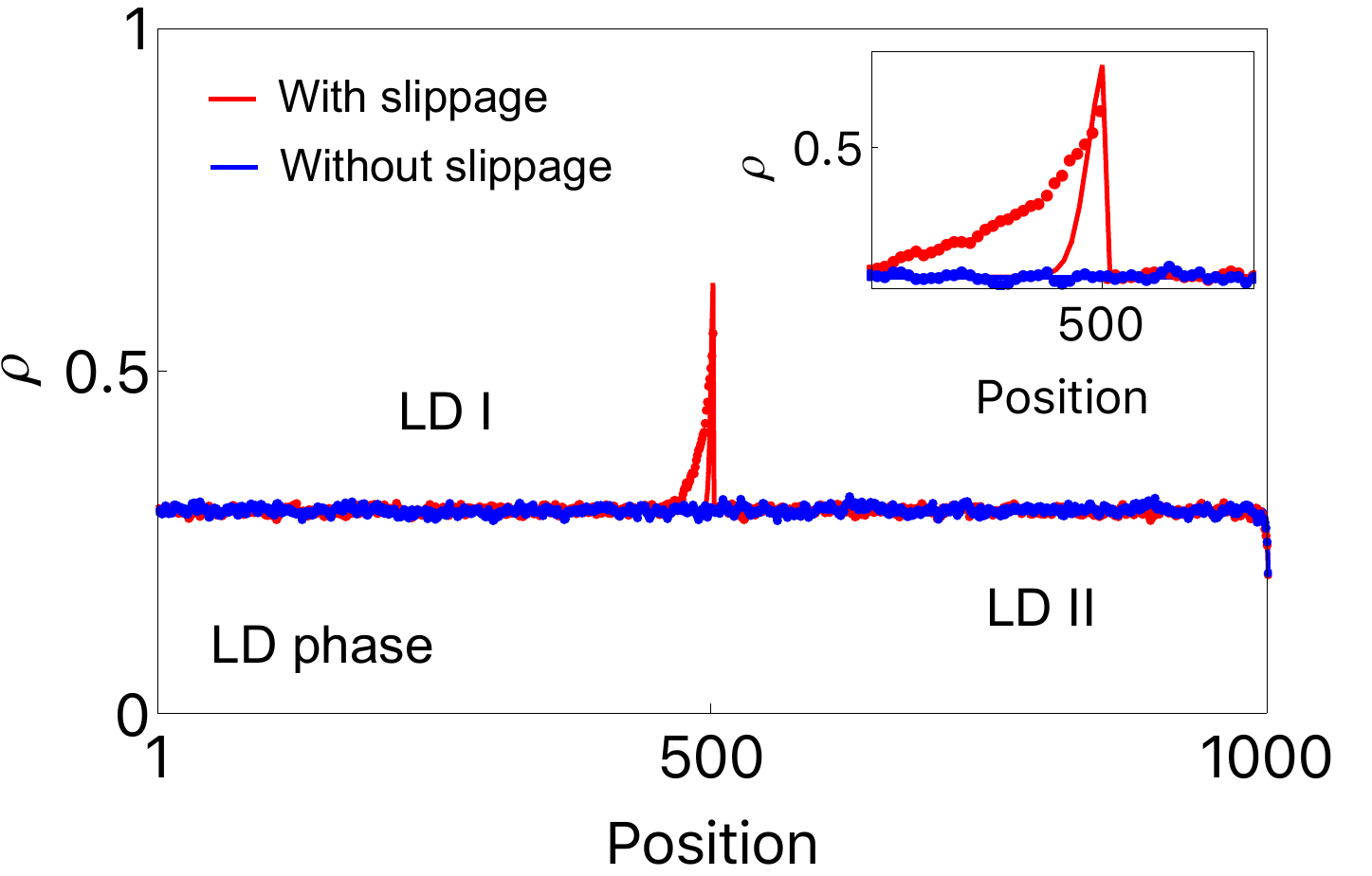}\\[0.02cm]
(b)\\[0.02cm]
\includegraphics[angle=0,width=0.9\columnwidth]{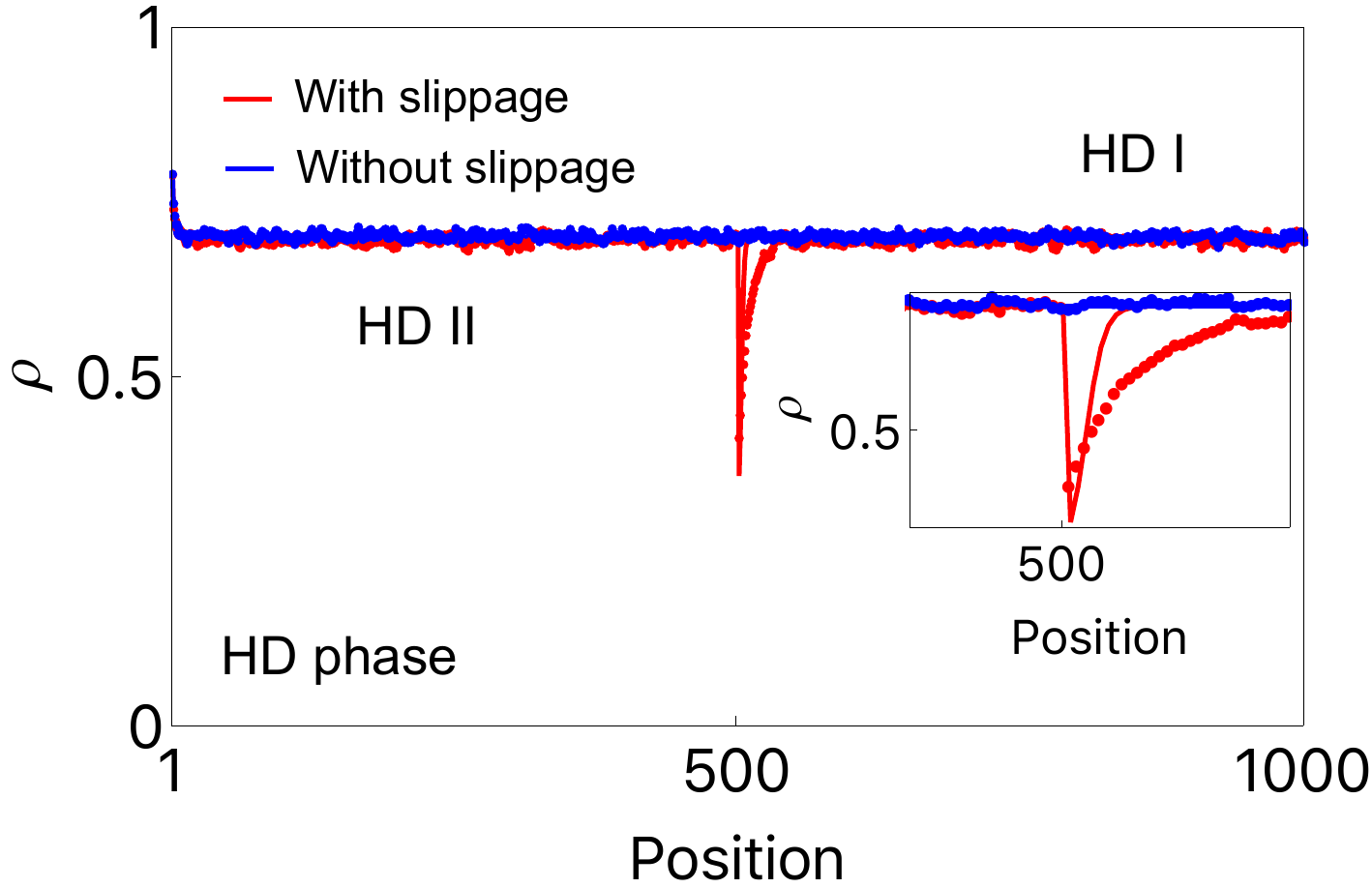}\\[0.02cm]
(c)\\[0.02cm]
\includegraphics[angle=0,width=0.9\columnwidth]{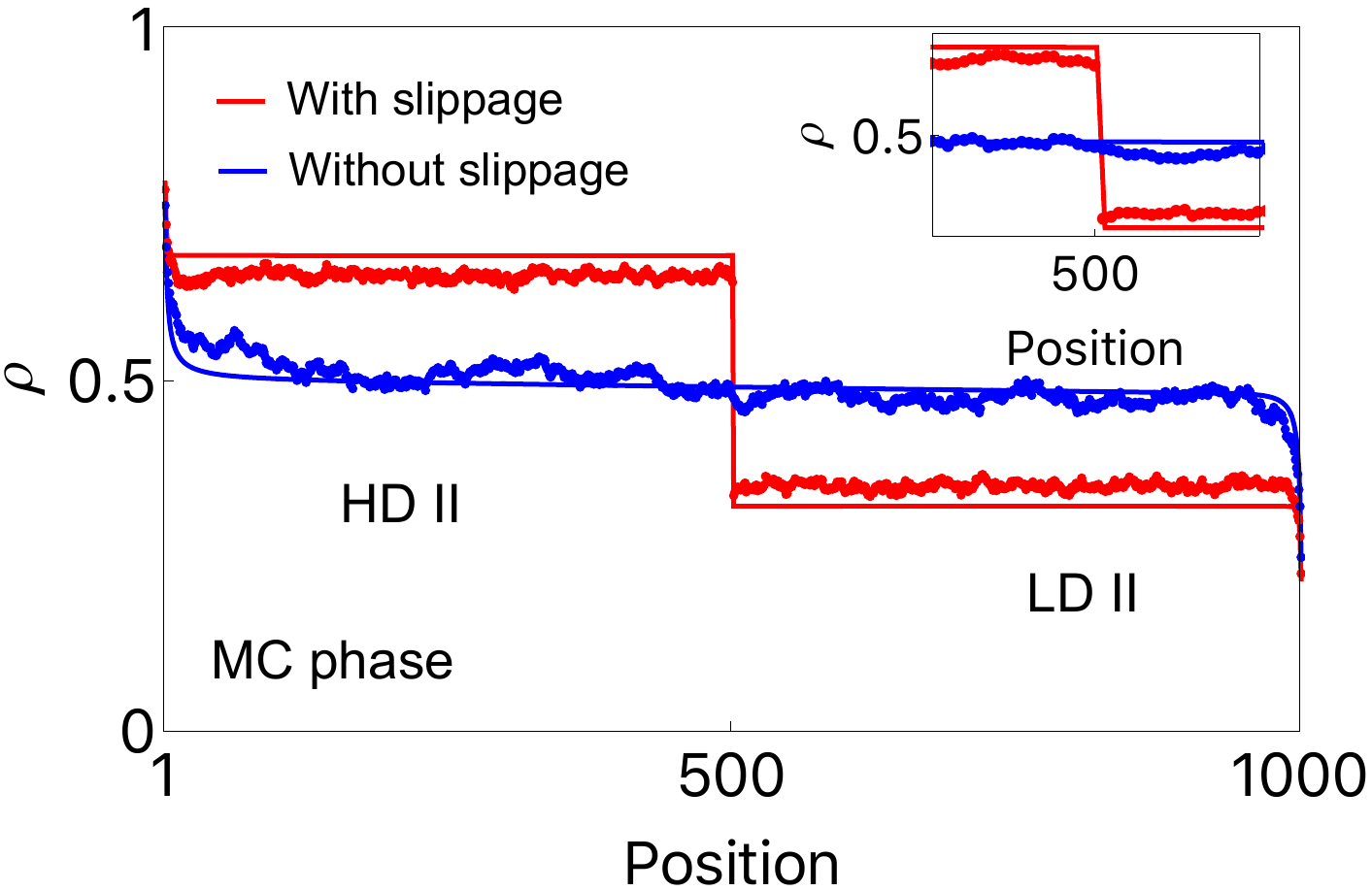}\\[0.02cm]
\end{center}
\caption{{\it Steady state} density profile of RNAPs traffic for (a) LD phase (b) HD phase and (c) MC phase. For LD: $\alpha=9~s^{-1}, \beta=30~s^{-1}$, HD: $\alpha=30~s^{-1}, \beta=9~s^{-1}$ and MC: $\alpha=30~s^{-1}, \beta=30~s^{-1}$. All other parameters remain fixed at values  $q=q_0=30~s^{-1}$, $q_{+1}=q_{-1}=20~s^{-1}$, $q_{+2}=10~s^{-1}$, $b_{1}=1000~s^{-1}$, $b_{2}=10~s^{-1}$ and $f_{1}=1~s^{-1}$. Lines correspond to MF theory and points have been obtained from  Monte Carlo simulations. }
\label{fig-result_densityprofile}
\end{figure}

In Fig.\ref{fig-result_meantime}, we have plotted {\it steady state mean time} ($T_{ss}$) as a function of $b_{1}$, for several different values of $b_{2}$. the trend of variations of the curves are very similar to those in 
Fig.\ref{fig-result_FPT}(a) except for the magnitude of the average times which in Fig. \ref{fig-result_meantime} is much longer because of the traffic congestion. The deviation of the theoretical predictions from the numerical simulations data in Fig. \ref{fig-result_meantime} is mainly because of traffic congestion. In maximal current (MC) phase, mean field (MF) results are known to deviate significantly from simulation results.

{\it Steady state} density profiles of the RNAP traffic in different phases are shown in Fig.\ref{fig-result_densityprofile}. We have taken the slippage site to be located at the mid-point of the track ($i = L/2$). This site can be regarded as a defect site on the homogeneous lattice. From the density profile plots, it is clear that the system behaves like a combination of two lattices (or two TASEPs). The TASEP 1 starts at site $i=1$, where particles can attach with attachment rate $\alpha$, and ends at site $i=L/2$ from where particles can detach with detachment rate $\beta_{eff}$; in between $i=1$ and $i=L/2$, particles can hop forward with rate $q$.  Similarly, the TASEP 2 starts at site $i=L/2+1$, where particles can attach with attachment rate $\alpha_{eff}$, and ends at site $i=L$ from where particles can detach with detachment rate $\beta$, in between the particles can hop forward with rate $q$. 

The LD and HD phases of the TASEP are further divided into the respective sub-phases LDI, LDII and HDI, HDII which are characterized by the nature of the decay of the density profiles in the boundary layers. In LD I phase of the TASEP the density profile at right boundary has positive curvature whereas in the LD II phase of the TASEP the density profile at right boundary has negative curvature. Similarly, in HD I phase of the TASEP the density profile at left boundary has negative curvature whereas in the HD II phase of the TASEP the density profile at left boundary has positive curvature. 

For the TASEP without slippage site, LD phase exists between $\alpha<q/2<\beta$. With slippage site, the TASEP 1 reaches LD I phase for $\alpha<q/2, \beta_{eff}<q/2$ (as the relation between the $\alpha$ and $\beta_{eff}$ does not matter) and the TASEP 2 reaches in LD II phase for $\alpha_{eff}<q/2<\beta$ (see in Fig.\ref{fig-result_densityprofile}(a)). Similarly for the TASEP without slippage site, HD phase exists between $\alpha>q/2>\beta$. With slippage site, the TASEP 1 reaches HD II phase for $\alpha>q/2>\beta_{eff}$ and the TASEP 2 reaches in HD I phase for $\alpha_{eff}<q/2, \beta<q/2$ (as the relation between the $\alpha_{eff}$ and $\beta$ does not matter) (see in Fig.\ref{fig-result_densityprofile}(b)). Similarly, for the TASEP without slippage site, MC phase exists between $\alpha>q/2, \beta>q/2$. With slippage site, the MC phase disappears and the TASEP 1 reaches HD II phase for $\alpha>q/2>\beta_{eff}$ and the TASEP 2 reaches LD II phase for $\alpha_{eff}<q/2<\beta$ (see in Fig.\ref{fig-result_densityprofile}(c)). 

\section{Summary and conclusion}

Motivated by the biological phenomenon of TS \cite{atkins10} in RNAP traffic, we have developed a stochastic kinetic model based on the TASEP where a special lattice site is treated as an unusual  `defect'. The state of each particle, which represents a RNAP, is denoted by two integer indices. The first index denotes its position on the lattice while the second expresses the extra length of the associated RNA transcript because of TS. 

In the first part of this paper, we have derived an exact analytical expression for the mean time taken by a single RNAP to traverse the defect site, in the absence of steric hindrance from any other RNAP. This mean time is extracted from the corresponding probability density distribution that we have derived here using the formalisms of first-passage time. The exact analytical expressions that we report reflect important statistical properties that characterize the passage of a single RNAP across the defect site while motoring along its DNA track. 

In the second part of this paper, we have investigated the interplay of TS at the defect site and RNAP traffic on the lattice where the RNAP traffic has been modelled as a TASEP. We have presented multiple evidences to establish increase in the number of TS events suffered by a RNAP while dwelling at the defect site for longer duration because of the traffic cogestion. We have also indicated how the TS process affects the flux of the RNAP traffic. 
We have found good agreement between our theoretical predictions, based on an approximate analysis of the TASEP model and the corresponding data obtained by carrying out Monte Carlo simulations of the same model. 

Our model is very general. However, our analytical treatment of the model is based on the assumption of homogeneity of the DNA sequence. In the realistic case, the transcription rate for every DNA nucleotide will vary depending on the identity of the nucleotide on the template and the concentration of the corresponding free monomers available in the surrounding medium. For quantitative predictions, that can be compared with experimental data, inhomogeneous sequence have to be considered. But, in those cases the analysis can be carried out only numerically because analytical treatment would be very difficult (if not impossible).

In spite of the simplicity of the special case of the model and the approximations made in its analytical treatment, the theoretically predicted probability distribution of the longer and shorter transcripts qualitatively matches the experimental data \cite{olspert15,olspert16} obtained through advanced sequencing technologies  \cite{beverly18}. 
However, for detailed quantitative predictions for specific systems, numerical values of the slippage rates would be required; but, at present, estimates of these rates are not available in the literature. 
We hope that our theoretical predictions will encourage more experimental studies.

The work reported here is of interdisciplinary nature. It has been motivated by a biological phenomenon,namely TS. The model proposed for studying this phenomenon is an extension of one of the simplest mathematical models of a system of interacting self-driven particles, namely the TASEP. The model has been analyzed from the perspective of non-equilibrium statistical physics. A single RNAP operates as a ``tape-copying Turing machine'' \cite{bennett82,mooney98,sharma12b}; a Turing machine \cite{feynman} is an idealized device conceptualized for abstract `computation' \cite{minsky67}. Therefore, the general theoretical framework developed here may be of interest also in the theory of computation.

\section*{Acknowledgements}
Work of one of the authors (DC) has been supported by a J.C. Bose National Fellowship from SERB (India) and a short-term Invitational Fellowship from JSPS (Japan). DC thanks KN for the invitation and hospitality at the University of Tokyo where part of the work was completed. DC also thanks the Visitors Program of the Max-Planck Institute for the Physics of Complex Systems for hospitality in Dresden when the revision of this manuscript was carried out. SP's visit to IIT Kanpur has been supported by an INSPIRE undergraduate scholarship from DST (India).


\end{document}